\documentclass[fleqn,usenatbib]{mnras}
\usepackage{newtxtext,newtxmath}

\usepackage[T1]{fontenc}
\usepackage{ae,aecompl}

\usepackage{graphicx}	
\usepackage{amsmath}	
\usepackage{caption}
\usepackage{subcaption}
\title[$R$-matrix and DW opacities for Fe~XVII]{Quantitative comparison of opacities calculated using the $R$-matrix and Distorted-Wave methods: Fe~XVII}

\author[F. Delahaye et al.]{
F. Delahaye$^{1}$\thanks{E-mail:franck.delahaye@observatoiredeparis.psl.eu},
C. P. Ballance$^{2}$,
R. T. Smyth$^{2}$
and N. R. Badnell$^{3}$
\\
$^{1}$Observatoire de Paris-LERMA, PSL Research University, Sorbonne Universit\'e, CNRS UMR 8112, F-75014 Paris, France\\
$^{2}$CTAMOP, School of Mathematics and Physics, Queen's University, Belfast BT7 1NN, UK\\
$^{3}$Department of Physics, University of Strathclyde, Glasgow G4 0NG, UK
}

\date{Accepted XXX. Received YYY; in original form ZZZ}

\pubyear{2020}

\begin{document}
\label{firstpage}
\pagerange{\pageref{firstpage}--\pageref{lastpage}}
\maketitle

\begin{abstract}
We present here a detailed calculation of opacities for Fe~XVII at the physical conditions corresponding to the base of the Solar convection zone. Many ingredients are involved in the calculation of opacities. We review the impact of each ingredient on the final monochromatic and mean opacities (Rosseland and Planck). The necessary atomic data were calculated with the $R$-matrix and the 
distorted-wave (DW) methods. We study the effect of broadening, of resolution, of the extent of configuration sets and of configuration interaction to understand the differences between several theoretical predictions as well as the existing large disagreement with measurements. New Dirac $R$-matrix calculations including all configurations up to the $n=$ 4, 5 and $6$ complexes have been performed as well as corresponding Breit--Pauli DW calculations. The DW calculations have been extended to include autoionizing initial levels. A quantitative contrast is made between comparable DW and $R$-matrix models. We have reached self-convergence with $n=6$ $R$-matrix and DW calculations. Populations in autoionizing initial levels contribute significantly to the opacities and should not be neglected. The $R$-matrix and DW results are consistent under the similar treatment of resonance broadening. The comparison with the experiment shows a persistent difference in the continuum while the filling of the windows shows some improvement. The present study defines our path to the next generation of opacities and opacity tables for stellar modeling. 

\end{abstract}

\begin{keywords}
Opacity -- $R$-matrix -- Distorted-Wave -- Solar model
\end{keywords}



\section{Introduction}

Opacities are key ingredients in any domain where radiative transfer is important. More specifically, Rosseland mean
opacities play an essential role in stellar modelling. They characterize the interaction between the photons 
produced in the centre of stars and the surrounding plasma up to the surface of the stars. Two of the main
historical providers of such quantities for stellar modeling are Forrest Rogers and Carlos Iglesias at the
Lawrence Livermore National Laboratory (LLNL) (project referred to as OPAL  \cite{OPAL1992}) and the Opacity Project 
(OP, \citep{ADOC_II}). Despite the fact that these two independent projects are based on largely different
physical frameworks, the agreement between the OPAL  \citep{OPAL1992} and the upgraded OP opacities
(\citep{Badnell2005}, hereafter OP2005) remains satisfactory for solar interior conditions and they both
originally accurately reproduced the helioseismic measurements. Those two datasets are widely used and they 
are references in stellar astrophysics (OPAL: https://opalopacity.llnl.gov/, OP: http://opacity-cs.obspm.fr/opacity/index.html, \cite{Delahaye2016}) . The good agreement between OPAL and OP is corroborated by more recent
efforts such as OPAS from the CEA (\citep{Blancard2012}; \citep{Mondet2015}) and LEDCOP (now ATOMIC) from the 
Los Alamos National Laboratory \citep{Colgan2013a,Colgan2013b, Colgan2015, Colgan2016}.
However, a revision of the Solar composition,  by Asplund and collaborators 
(\citep{Asplund2004, Asplund2005, Asplund2009}) resulted in a reduction of the content of C, N, O and Ne 
by $30-40\%$. This has significantly  degraded the much coveted agreement between stellar theory and 
helioseismic measurements. For around 15 years, in an effort to solve this Solar abundance problem, atomic
theorists and experimentalists have been working hard to improve the quality of the opacities. Meanwhile, 
a new discrepancy appeared in 2015 with the measurement at Sandia National Laboratory of the Fe opacity  at
$180$~eV  and $N_e=3.1\times 10^{22}~$Scm$^{-3}$, which is in line with conditions corresponding to 
the base of the Solar convection zone: $T_e\approx 2.15\times 10^6$~K and $N_e\approx 3.1\times 10^{22}~$cm$^{-3}$. 
Indeed, while all the theoretical calculations agree reasonably well, the experiment measured an opacity a 
factor of 2 higher than all calculations and exhibits large differences (filled windows, higher continuum) 
at certain photon energies \citep{Bailey2015, Nagayama2019}. 

In order to shed light on this long standing problem of the Solar abundances and opacities, a detailed study of each
component contributing to the determination of the opacity is required. Such a comparison is complex as 
each different step and input choice affects the final results in many, and quite often correlated, ways. Our goal 
is to quantify these effects by 
comparing the two main approaches (distorted-wave (DW) and $R$-matrix) employed  by the theoretical groups who calculate fundamental atomic rates, as well as all the secondary derived values of the calculations such as populations, broadening, convergence and resolution.

In the following two sections we summarize the $R$-matrix and DW methods and we provide detailed 
descriptions of the Fe XVII models used. In Section 4 we present the results for Fe~XVII opacities and comment
on the effect of broadening, the convergence configuration expansions in both the initial and final levels 
of photo-absorption, etc. We provide a summary and our conclusions in the last section.

\section{Methods:R-matrix-DARC}

The parallel version of the DARC (Dirac Atomic $R$-matrix Codes) \cite{Norrington1987,Ballance2006} was employed to calculate the photo-absorption cross-sections from several hundred initial 
states of Fe XVII to provide the bound-bound and bound-free components of a Rosseland mean opacity, subsequently calculated with the Opacity Project suite of codes.
 Three models were employed to quantify the self-consistent convergence of opacities within the $R$-matrix theory before comparison with distorted-wave calculations.

Usually, $R$-matrix photoionization calculations are initiated by generating a set of orbitals 
optimised on the levels of the residual (F-like Fe) ion. To achieve this we employed a modified version
of the GRASP$^0$ code \cite{Grant1980} that implements the Multi-Configuration-Dirac-Fock (MCDF) method to provide the orbital radial wavefunctions upon a numerical grid, that may be  
subsequently employed by the following DARC calculation.
Our initial calculation (Model A) includes the $2s2p^5nl$ and $2s^22p^4nl$  configurations in which
$nl=2p, 3s, 3p, 3d, 4s, 4p, 4d, 4f$.
Model B extends Model A to $nl=5s, 5p, 5d, 5f, 5g$.
Model C further extends Model B by the inclusion of additional $nl=6s, 6p, 6d, 6f, 6g$ orbitals.  

The transition from Model B to Model C increases the number of levels included in 
the close-coupling expansion from 407 levels to 638 levels. This also expands the $R$-matrix box from 4.2 Bohr radii (Model B) to 7.2 Bohr radii (Model C). Theoretically, these changes seem minimal, but they lead to an increase in the Hamiltonian size from approximately 75,000 $\times$ 75,000 to matrices close to 100,000 $\times$ 100,000, with an associated $N^3$ scaling in the computational effort when diagonalising the Hamiltonian for the initial and final state eigenvectors. The $R$-matrix method 
inherently includes all the Rydberg series attached to every target state of the residual ion 
which is dependent on the resolution of the energy mesh employed.

  The parallel treatment of the outer region benefits from GPU (Graphical Processing Units) enabled codes for determining initial boundstate levels and for the bound-free photo-absorption. For Model 
  B, all the initial states (up to $n=5$) are truly bound  whereas for Model C
  some high lying ($n=6$) initial states associated the the $2s2p^6$ core are above the Fe XVII ionization 
  limit. 
  Without the GPU enabled codes, it would be difficult to achieve the energy resolution 
  described in the following sections within a feasible time. Model C has several hundred initial states, each with a fine resolution, and post-processing by the opacity codes requires significant amounts of time.

\section{Methods: AUTOSTRUCTURE (DISTORTED-WAVE)}
We use the program AUTOSTRUCTURE (AS) \citep{Badnell2011} within a perturbative DW approximation to calculate the required photoabsorption data in a multi-configuration intermediate coupling Breit–-Pauli model. Photoabsorption which leads to the direct ejection of an electron into the continuum (photoionization) is calculated independently of photoexcitation to another discrete atomic state. This final discrete state may be truly bound or lie
above the ionization limit. We calculate the Auger width as well for the latter case. The total 
width of the upper state is then an independent sum of the natural (radiative) width, 
the Auger width, and the collisional broadening width. See \cite{BadnellSeaton2003} for details. 
This enables us to make detailed comparisons with $R$-matrix results which are intrinsically Auger-broadened. The multireference (MR) configuration sets for the Fe~$^{16+}$ photon target and the Fe~$^{17+}$ residual ion (electron target) are simple to list in terms of promotion rules. (The $1s^2$ is kept closed through-out and so is omitted in the following descriptions.) We consider first (MR1)

Fe$^{16+}$: $ 2s^22p^6$ 
\newline
and we allowed for single and double electron promotions (of $2s$  and $2p$) up to some $n=n_{\rm max}$.

Fe$^{17+}$: $2s^22p^5, 2s2p^6$
\newline
where we allowed for single $n=2$ electron promotions up to the same value of
$n_{\rm max}$.

Due to differences in the structure of the F-like electron target and corresponding Ne-like photon target used by DW and $R$-matrix, the number of true bound levels differs even when the exact same configuration expansions are used.

The contribution from $n > n_{\rm max}$ is determined by extrapolation of the $n=n_{\rm max}$
photoionization data to higher $n$. The extrapolated contribution to photoexcitation is determined by extrapolating threshold or edge photoionization data (including the newly extrapolated) down below threshold. We allow for configuration mixing between all configurations within an ionization stage. This
is in contrast with the updated Opacity Project work which restricted the mixing to within the $n=2$ core complex.
We calculated photoabsorption data for all possible initial Boltzmann populated levels arising from this MR1 set. This includes autoionizing levels. This is in contrast with the updated Opacity Project work which restricted the initial levels of the AS calculation to true bound so as to match with those of the original Opacity
Project $R$-matrix calculations.

We note that the above MR1 set has at most two L-shell vacancies in the final state. We considered also a supplementary MR2 set:

Fe$^{16+}$: $ 2s^22p^43s^2$, $2s2p^53s^2$, $2p^63s^2$
\newline
and we allowed for single and double electron promotions from $3s^2$ as well as, for PE only, single promotions from $2s$ \& $2p$, up to $n_{\rm max}$.

Fe$^{17+}$: $ 2s^22p^33s^2$, $2s2p^43s^2$, $2p^53s^2$
\newline
where we allowed for single and double electron promotions from
$3s^2$ up to the same value of $n_{\rm max}$,
for PI only. We did not calculate Auger widths in this instance since we were not comparing this data with $R$-matrix and collisional broadening dominates over the Auger width in general.
This MR2 set enables us to determine the inner-shell contribution to opacity 
from two- to three-L-shell vacancy transitions.
We constrain it to give only this contribution so that it can be added to that from MR1. We denote the combined dataset MR1+MR2.

Given the above set-up, all required energy levels, radiative and autoionization rates, and photoionization cross sections were calculated using the AUTOSTRUCTURE code \citep{Badnell2011} in the manner described by \cite{BadnellSeaton2003}. We carried-out calculations for $n_{\rm max}=5$ and $n_{\rm max}=6$ so as to test the convergence of the extrapolations. From here-on we drop the subscript and just refer to these calculations as $n=5$ and $n=6$, along with the MR set MR1 and/or MR2 as necessary.

\section{Opacities}

The calculation of opacities requires several steps. It is necessary i) to determine the ionic state of the absorbing medium as well as the population of the atomic levels of its constituents, ii) to calculate the absorbing coefficients for different processes (photo-excitation, photoionization), and iii) to include plasma effects (the ions cannot be considered as isolated). Such calculations, despite the computer power available today, still require approximations and compromises. Accuracy versus completeness is still at play nowadays as it was 30 years ago: the atomic systems considered may be much larger, but they cannot be infinite. Several groups (LLNL-OPAL, CEA-OPAS, LANL-ATOMIC, The Opacity Project) are involved in such calculations, using their own methods and recipes at each step. Nonetheless, the various modern Rosseland mean opacities for a solar mixture at the conditions found in the Sun agree to within 4\%. However, when we compare the monochromatic opacities for individual elements, the differences are much larger. The mixture and the averaging over the frequencies introduce some cancellation effects. Detailed comparisons are thus required for a single element at different conditions in order to disentangle the different effects from the different ingredients, and more importantly to characterize the signature of each approximation on the final results. While comparisons with the few available experiments appear necessary, a word of caution is relevant here: in such comparisons, experimental uncertainties must be clearly stated.   

In the present section we consider all the different stages of an opacity calculation and try to identify the uncertainties and their impact on the final result. We first consider the $R$-matrix calculation of the photoionization cross sections and transitions probabilities, looking at the effect of the configuration expansion, resolution and extrapolation. Then we present the results for the DW approach and compare them with the $R$-matrix ones.
All the calculations are level-resolved. Finally, we compare the present results with other models OP2005 \citep{Badnell2005}, SNAKP \citep{Nahar2016a,Pradhan18} and OPAS \citep{Blancard2012} and SCRAM \citep{Hansen2007} as well as with the experimental results from SANDIA \citep{Bailey2015,Nagayama2019}. All the opacity results presented here are for the physical conditions corresponding to those in the SANDIA experiment 
($T_e=2.1\times 10^6$~K and $N_e=3.1\times 10^{22}$~cm$^{-3}$).

\subsection{R-matrix results}

\subsubsection{Convergence: Model A ($n=4$), Model B ($n=5$) and Model C ($n=6$)}

One of the problems in opacity calculations comes from the difficulty to test the convergence of one isolated ion stage. Indeed, despite the ever growing computer power, increasing the number of configurations is a challenge. 
In the present calculation we have run 3 sets of configurations up to the complex $n=4$, $n=5$ and $n=6$ corresponding to Models A, B and C, described previously. These 
relativistic calculations include respectively 267, 407 and 638 levels in the close-coupling expansion of the residual ion target. 
One might expect an automatic increase in the resulting opacities since more open channels as well as more resonance structure are associated with an ever increasing target expansion. As we can see in Figure 1, the monochromatic opacities do exhibit a net 
effect. There are more resonances appearing when we include more configurations with higher $n$-shells, the continuum increases at low photon energy but 
seems to be lowered at high energy in spite of increasing numbers of open channels. In fact, one has to remember that increasing the size of the target and hence initial levels taken into account also means 
a redistribution of the population of the contributing levels, which corresponds to a dilution of the 
population. The strong contributors in the $n=4$ expansion will see their population reduced as we extend the 
calculation to $n=5$ and then to $n=6$. 
For the Rosseland mean opacity, this effect of dilution, from $n=4$ to $n=5$ is largely compensated by the extra 
contribution to the photo-absorption while it is not compensated going from $n=5$ to $n=6$. The Rosseland mean opacity $(\kappa_R)$ of Fe~XVII is the $\kappa_R=1.2451\times 10^{-3}$ for $n=4$, $\kappa_R=1.2888\times 10^{-3}$ for $n=5$ 
and $\kappa_R=1.2834\times 10^{-3}$ for $n=6$. 
While the impact on the Rosseland mean is mild, changes in the Planck mean $(\kappa_P)$ are more significant,
$\kappa_P=1.3218\times 10^{-2}$ for $n=4$,  $\kappa_P=1.1049\times 10^{-2}$ for $n=5$ and  
$\kappa_P=1.3489\times 10^{-2}$ for $n=6$. It represents a difference of $\sim 2\%$ and $\sim 18\%$ for the Planck mean between $n=6$ and $n=4$ and $n=6$ and $n=5$, respectively, while the change in the Rosseland mean is below $4\%$ and $1\%$, respectively. The differences in the Planck mean are surprising but a word of caution is relevant here as the 3 sets of calculations are not at the same resolution. We note that the $n=4$ calculation employs $5000$ individual energy points, $n=5$ uses $40000$ points and $n=6$ uses $80000$ points. This may partly explain the differences in the Planck mean as we will see in the next subsection. All these results for the mean opacities were obtained with an opacity sampling of $10^4$ points in energy. Higher resolution ($10^5$) has also been used and it will be discussed later.

\begin{figure}
 \includegraphics[width=\columnwidth]{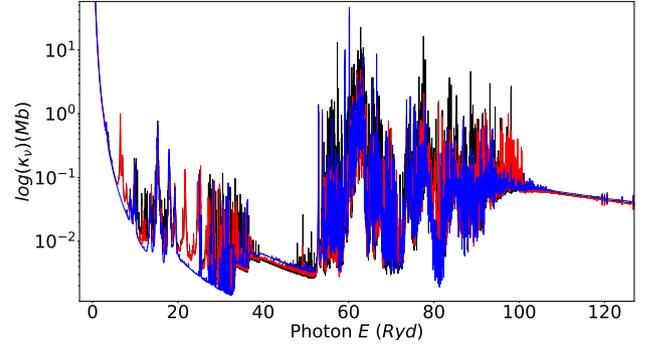}
\caption{$R$-matrix monochromatic opacities for Fe~XVII for configurations up to $n=4$ (blue), $n=5$ (red) and $n=6$ (black). (Models A, B and C.)}
    \label{fig:Fig1}
\end{figure}

\subsubsection{Resolution: Model C at 10k, 20k, 40k and 80k points}

In the $R$-matrix calculation of photoionization cross sections, the resolution plays a crucial role since it will directly impact the capability to resolve all the resonances present in the calculation. It will have a greater impact on the Planck mean opacity since it is a classical mean which depends directly on the height of the resonances.  On the contrary the Rosseland mean being a harmonic mean, depends essentially on the lows of the cross sections hence should not be too sensitive to the resolution since in the present work we do not apply any broadening of such resonances. We expect a larger effect from the broadening of the resonances because it will automatically fill the windows between resonances and may even raise the continuum over a large range of energies due to the overlapping of the broadened resonances. However, if the resonances are numerous, even without broadening, the resolution could be an issue.

We can see this effect in Figure 2 where cross sections for Model C with different resolutions are presented. 
The peak of resonances goes up and down depending on the resolution. From 10k to 80k the difference in $\kappa_P$
reaches $6\%$. Then the differences oscillate between $+2\%$ and $-2\%$ for 20k and 40k when compared to 80k. 
The Rosseland mean does not change by more than $0.2\%$ between the different resolutions, as expected.

\begin{figure}
 \includegraphics[width=\columnwidth]{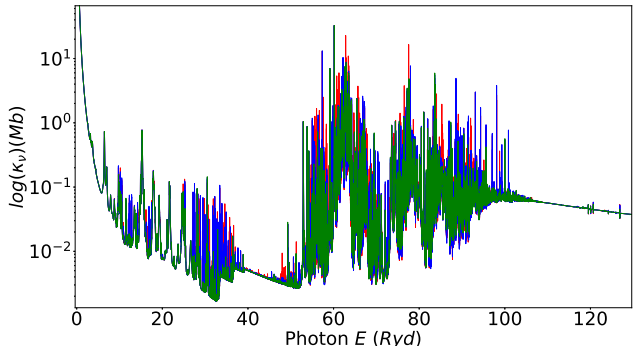}
\caption{$R$-matrix monochromatic opacities for Fe~XVII (model C) using high resolution cross sections (80k points - black), medium resolution cross sections (40k points - red), an intermediate resolution and a low resolution (respectively 20k-blue and 10k-green points). The difference for $\kappa_P$ is $\sim 2\%$ between 20k/40k and 80k but reaches $\sim 5\%$ for 10k. It remains below $\sim 0.2\%$ for $\kappa_{R}$ regardless of the resolution.}
    \label{fig:Fig2}
\end{figure}

In the OP opacity calculation there is also another aspect concerning the resolution. This is the sampling. Indeed, 
to reduce the number of energy points used in monochromatic opacities, we use an adaptive sampling in order to 
emphasize the energy region where the contribution is maximum \cite[see][for details]{Seaton1994}. 

As mentioned in \cite{Seaton1994} the means do not change by more than 0.1\% when the sample is done 
with $10^5$ points or more and up to 2\% for $10^4$ points. We found this sampling independent of the 
original resolution in the present cases (for all the different energy resolution in the cross sections 
from 10k to 80k). We adopted the fine energy sampling ($10^5$ points) for the present study.

\subsubsection{Additional bound-bound transitions.}
In order to improve the calculation it is important to try to take into account all the possible contributions to the
absorption. Using the $R$-matrix method, the photoionization calculation is the most demanding. Hence we cannot extend 
much more above the complex $n=6$. However, the contribution from the bound-bound transitions being easier to generate,
we can extend it to higher levels. We added the bound-bound contributions from levels up to $n=9$. This correspond to ~3.5 times more bound-bound transitions (from ~4000 to ~13000 lines).
While we might expect some increase in opacities from these extra radiative transitions, as explained
above, it can be counter-balanced by the dilution of the population. 
In Figure 3, we see many lines (in blue compared to the red curve) added in the low energy ranges, and a clear lowering of the continuum at all energies as well as a reduced height of the peaks.  In order to highlight the dilution effect we suppressed the broadening of the bound-bound transition for both sets.
When broadening is applied (see Figure 4) we see an increase of the continuum in the range [25 - 30 Ryd]. This is due to the broadening of the numerous lines in this region as seen in Figure 3. Otherwise we can distinguish the lowering of the continuum at all energies beside the region mentioned above, due to the dilution of the populations. The changes in the level populations also affect the  peaks everywhere. The net total effect is an increase of the Rosseland mean opacity and a decrease of the Planck mean. The Planck mean is reduced from $\kappa_P=1.3472\times 10^{-2}$  to $\kappa_P=1.2597\times 10^{-2}$ and 
the Rosseland mean increases from $\kappa_P=1.34764\times 10^{-3}$ to $\kappa_P=1.4559\times 10^{-3}$.

\begin{figure}
 \includegraphics[width=\columnwidth]{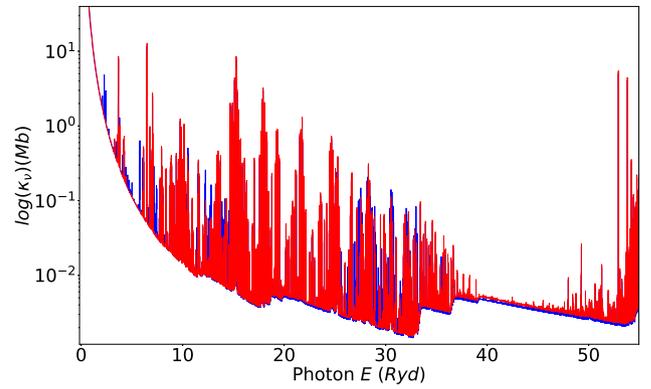}
\caption{$R$-matrix monochromatic opacities for Fe~XVII with bound-bound transitions up to: $n=6$ (red, corresponding to 172 starting levels) and up
to $n=9$ (blue, corresponding to 327 starting levels ). The bound-free contribution is the same, including levels up to $n=6$. The broadening of the bound-bound transition has been reduced (divided by 1000) in order to outline the presence of the extra lines and the net effect when broadened is presented in Figure 4.}
    \label{fig:Fig3}
\end{figure}

\begin{figure}
 \includegraphics[width=\columnwidth]{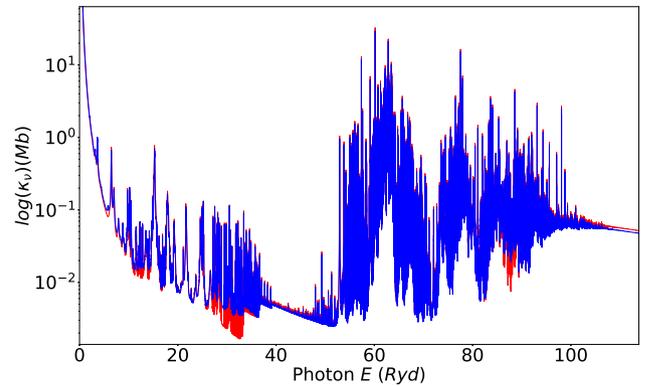}
\caption{$R$-matrix monochromatic opacities for Fe~XVII with bound-bound transitions up to: $n=6$ (red, corresponding to 172 starting levels) and up
to $n=9$ (blue, corresponding to 327 starting levels ). The bound-free contribution is the same, including levels up to $n=6$. The broadening of the extra bound-bound transitions induces a pseudo continuum at low energy where most of the extra lines are.}
    \label{fig:Fig4}
\end{figure}


\subsubsection{Extrapolation of the cross sections for starting levels of Rydberg series.}
Another contribution can easily be added. The absorption from starting levels belonging to a well identified Rydberg series can be extrapolated from the highest levels belonging to the same Rydberg series with calculated photoionization cross sections. The highest $n$-shell explicitly calculated was $n=6$, and therefore taking into account the change in the effective quantum number along each Rydberg sequence we were able to approximate small contributions from $n=7-9$ initial states from the $n=6$ results. 

In the present case we identified 28 Rydberg series among all the 327 levels up to n=9 and we have generated 102 extrapolated cross sections. The final total number of starting levels is 358 (some of the extrapolated cross sections are attached to existing levels belonging to the set of 327 up to the complex $n=9$). 
Among the levels for which new cross section have been produced via these extrapolation, most of them were already contributing to the bound-bound absorption described in the precedent section. These extrapolations bring more resonances and increase slightly the continuum, compensating sufficiently the extended dilution of the populations. Both mean opacities ($\kappa_P$ and $\kappa_R$) are affected and increased by such procedure by 2\% and 3\% respectively.

\subsubsection{Resonance broadening}

While the bound-bound transition broadening is completely treated in our calculation, this is not yet done for the resonances present in the bound-free cross sections. Indeed, unlike the case of DW, the inclusion of broadening into the $R$-matrix approach is not trivial. A difficulty comes in practice due to overlapping resonances arising from different series which, individually, require rather different collisional broadening.
The $R$-matrix results reported in \cite{Nahar2016a,Pradhan18} include a treatment of such broadening, but they give no details of their methodology. 
We discuss their results later in section 4.3.2.

\subsection{AS-DW results}
Historically, The Opacity Project has favored the $R$-matrix approach for the calculation of the atomic
data used in the opacity calculation since OP focuses on the quality of the atomic data. Of course
such a choice implied some assumed approximations in the treatment of plasma effects. In contrast, all other groups
favoured the description of the plasma itself at the cost of approximations in the atomic data
calculations. In the opacity calculations it is always necessary to build up a compromise between accuracy and
completeness of the atomic data sets and of the plasma effects. In 2005, an Opacity Project team have extended the validity of the OP opacities with the inclusion of inner-shell transitions \cite{Badnell2005} using the DW approximation. Hence, we combined both methods in order to improve our results for stellar applications.
The less complex DW method allows for larger configuration sets. Another important difference between the DW and the $R$-matrix approaches is the treatment of photoionization. There are two paths to
photoionize an atom/ion. One is a direct path with a photon with enough energy to detach directly an
electron and send it into the continuum. The second way consists in photo-exciting the atom/ion into
an autoionizing state which decays after, sending an electron into the continuum. This corresponds 
to the photo-excitation of an electron into a quasi-bound state followed by a decay which makes
another electron free via the Auger effect. This can be characterized by the two following equations:

\begin{equation} \label{eq1}
\begin{split}
X^{n+} + h\nu \rightarrow X^{(n+1)+} + e^{-}
\end{split}
\end{equation}

\begin{equation} \label{eq2}
\begin{split}
X^{n+} + h\nu \rightarrow X_{*}^{n+} \rightarrow  X^{(n+1)+} +  e^{-}
\end{split}
\end{equation}

These two processes are treated directly at once by the $R$-matrix method, including the
interaction between the two processes, while the DW approach treats the two processes separately. 
This has a direct impact on the shape (i.e. the Fano profile) of the resonances present in the photoionization cross sections. No such profiles are present in DW calculations and the cross sections are a superposition 
of a continuum and simple lines with a Voigt profile. A detailed comparison has been presented in \cite{Delahaye2013}.

All the present calculations were performed in intermediate coupling. The comparison with previous LS coupling results can be found in \cite{BadnellSeaton2003, Delahaye2016}. The effect of configuration interaction has been presented in \cite{Delahaye2016}.

\subsubsection{$n=5$ vs $n=6$}

Similarly to our $R$-matrix computations, we study the effect of the configuration sets. We have calculated the radiative data for two different sets of MR1 configurations, $n=5$ and $n=6$, as described in section 3. The first set corresponds to 157 true bound starting levels of Fe~XVII and 501 levels of the residual ion.  The second set includes 219 true bound starting levels and 779 levels in the residual ion. Passing from one set to the other, as we can see in Figure 5, where the $n=5$ set is in red and $n=6$ in black, the extra contribution as we did in the $R$-matrix results. For example, around $90$~Ryd, the extended calculation provides new absorption which fills-in the gap seen in the $n=5$ (red) results. We have to emphasize the fact that in the DW approach the lines from bound-bound and the resonance from bound-free processes are both broadened. It is the overlapping of these broadened lines that fill-in the gap, as we will detail in the next subsection. The extra levels in the $n=6$ calculation also generate a dilution effect due to the different level populations. We clearly see it in the energy range $[20-50$ Ryd$]$. The impact is much more pronounced in the DW approach than in the $R$-matrix treatment. This is also coming from the fact that in this comparison we took only into account the contribution of the true bound levels as starting levels. Extending the configuration set changes the structure of the ion and some levels near the ionization limit shift into the continuum as quasi bound states and disappear from the starting list of bound levels used to calculate the opacities. This reduces significantly the final bound-free contribution to the opacities despite the reduction of the dilution.

\begin{figure}
 \includegraphics[width=\columnwidth]{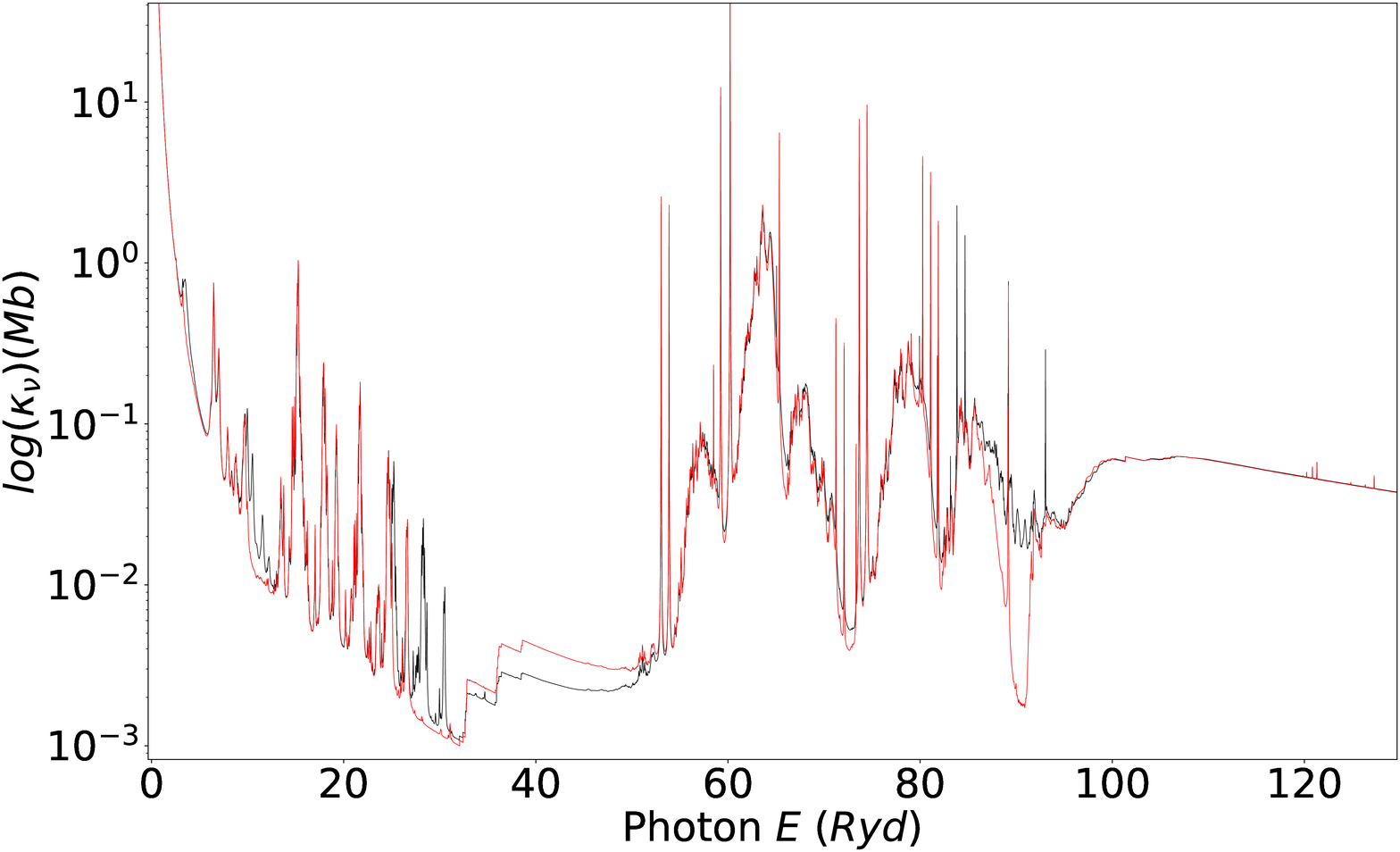}
\caption{Comparison between DW monochromatic opacities using two sets of MR1 configurations $n=5$ (red) and $n=6$ (black).}
    \label{fig:Fig5}
\end{figure}

We clearly see the importance of the addition of higher $n$ transitions. While the $R$-matrix method for example automatically includes the resonances from high-$n$ Rydberg series depending only on the resolution, it has to be done explicitly in the DW method. Since we cannot include an infinite number of configurations, we use an extrapolation scheme in order to take into account the contribution of the resonances converging to each edge of the photoionization cross sections.  As we did in the previous OP opacity release in 2005 \cite{Badnell2005}, we extrapolate the direct cross section below the edges so as to give a smooth transition of  the numerous broadened resonances converging towards the edge. As seen in Figure 6 the impact may be important in some cases. It fills in gaps over a large range of energy and has a strong effect on the Rosseland mean, especially when it covers the spectral range of maximum contribution (peak of the weighing function) which corresponds to $50$~Ryd in the present case (at $u=\frac{h\nu}{kT}\sim3.8$ with $T=180$ eV). This corresponds to an increase of $30\%$ in the present case ($n=5$, true bound starting levels only). As we expand the configuration set the effect is reduced, as expected and is $20\%$ in the $n=6$ case. But we cannot go much further since it would mean extending dramatically the configuration sets.

\begin{figure}
 \includegraphics[width=\columnwidth]{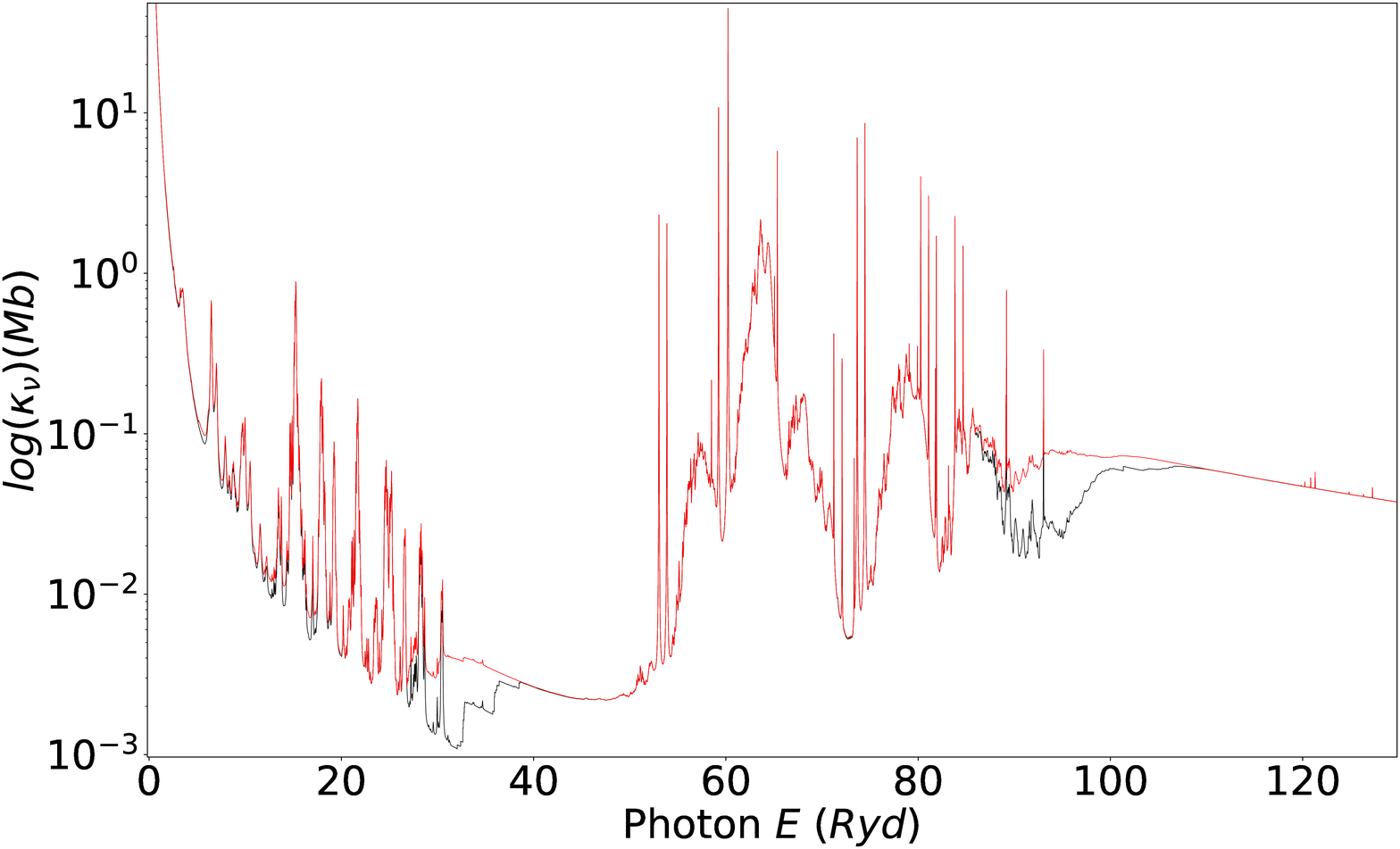}
\caption{Comparison between DW monochromatic opacities for MR1 $n=6$ without the extrapolation below the edge (black) and with extrapolation (red) to account for missing high-$n$ Rydberg resonances.}
    \label{fig:Fig6}
\end{figure}

When we compare the two sets of configurations $n=5$ and $n=6$ with the extrapolation (Fig. 7) including the extrapolation features, we see a closer agreement, especially in the high energy region. The remaining large discrepancies in the crucial $[20-50$ Ryd$]$ energy range come from the mismatch between the starting levels taken into account due to the differences between the two structures. This will be addressed in the next section. 

\begin{figure}
 \includegraphics[width=\columnwidth]{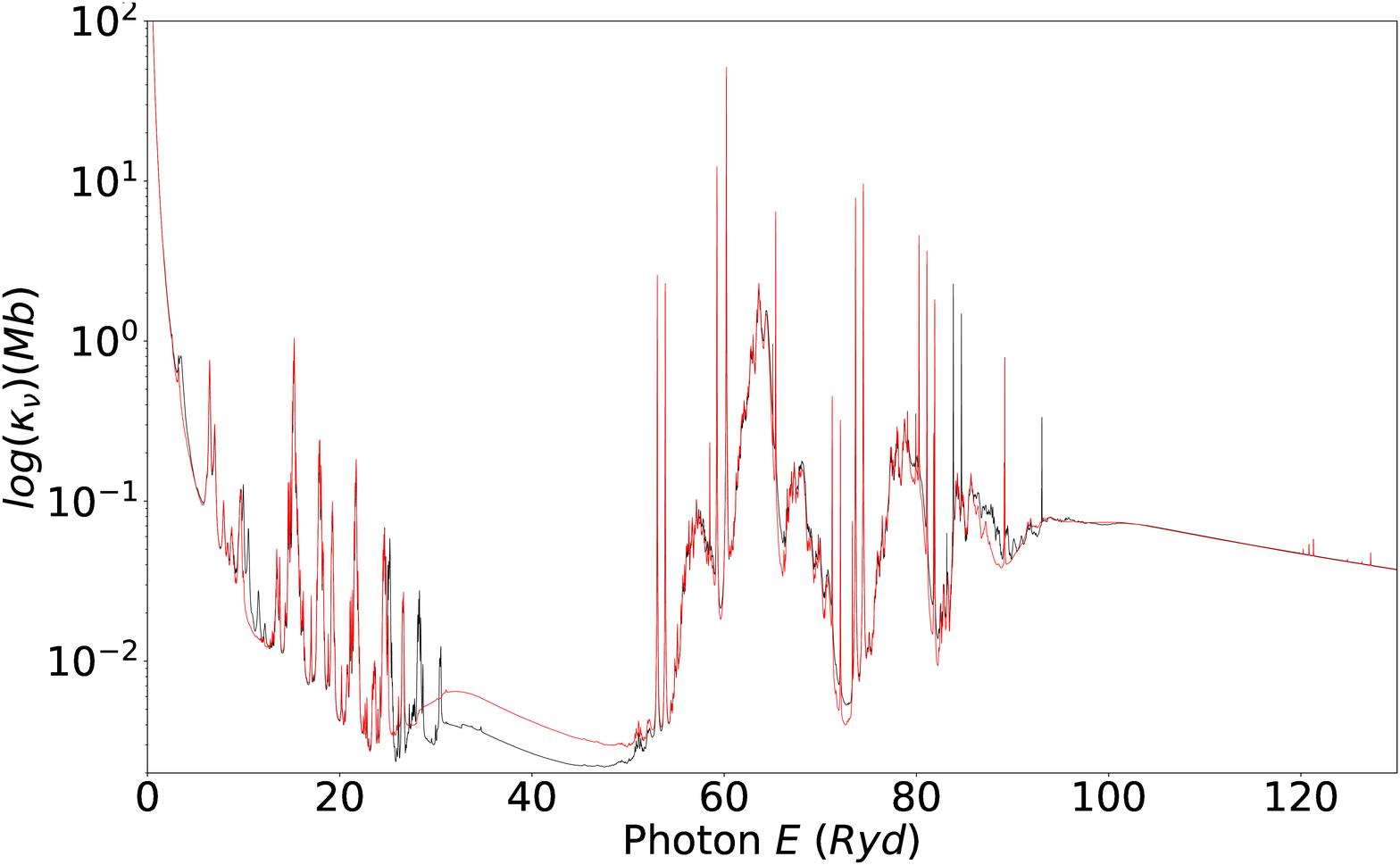}
\caption{Comparison between  DW monochromatic opacities for Fe~XVII with the MR1 configurations up to $n=5$ (red) and $n=6$ (black) with extrapolation below the edge for both (see text for details).}
    \label{fig:Fig7}
\end{figure}

\subsubsection{Extension: Inclusion of quasi-bound states and triple $L$-shell holes}

One advantage of the DW approach is that it allows for the inclusion of many more configuration and levels, even autoionizing levels as initial states. If their population are sufficiently large, then they will make a contribution in the bound-bound transition as well as in the photoionization. The $R$-matrix method takes into account only true bound levels. 
as starting levels. The net result is difficult to evaluate as the impact on the population is counterbalanced by the extra absorption.

\begin{figure}
 \includegraphics[width=\columnwidth]{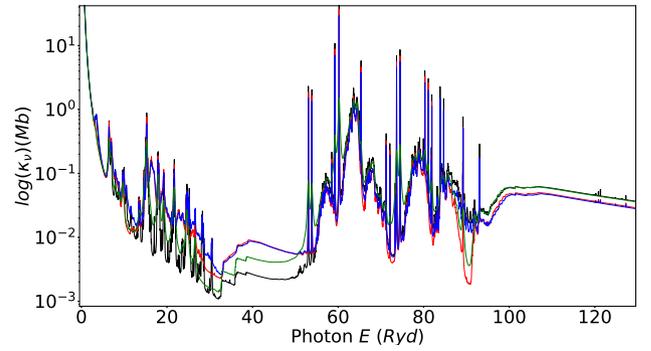}
\caption{Comparison between DW monochromatic opacities for MR1 configurations with $n=5$ only true bound starting levels (green), $n=5$ all starting levels (true bound plus autoionizing - red), $n=6$ only true bound starting levels (black), $n=6$ all starting levels (blue). For clarity the extrapolation below the edges has been omitted for all calculations here.}
    \label{fig:Fig8}
\end{figure}

In Figure 8, we have plotted the spectral cross sections of Fe~XVII for 4 different sets of configurations, without including the extrapolation of the high-$n$ Rydberg resonances for clarity. In blue and green we show the data presented above for MR1 $n=5$ and $n=6$ true-bound starting levels while in red ($n=5$) and black ($n=6$) we have included all starting levels (true bound plus autoionizing). While we had 157 and 219 starting levels for MR1 $n=5$ and $n=6$ when only true bound states were included now we have 19635 and 51179 starting levels respectively with the autoionizing starting levels. The extra contribution dominates strongly the dilution when compared with the calculation containing only true bound starting levels in the  $[20-50$ Ryd$]$ energy range but at high energy, the dilution is not fully compensated-for. Hence, this will increase the Rosseland mean opacity but it will not help to increase the continuum at high energy, as the experiment \citep{Bailey2015} tends to suggest. Between $n=5$ and $n=6$ (bound plus autoionizing) the increase on starting levels (more than twice as many levels) generates a dilution compensated-for by the added contribution to the bound-bound absorption and bound-free cross sections. 

We have extended the exercise to the inclusion of transitions with double-to-triple L-shell vacancies. One might have anticipated that their contribution would be minor since the
starting levels should have a small population. However, as we can see on Figure 9, their contribution at high energy is significant. First, we can see some windows being filled ($E \in [65,73$ eV$]$ and $E\in[81,92$ eV$]$). Second, at even higher photon energy, the continuum is significantly raised. The total impact on the Rosseland mean opacity of Fe~XVII is an increase of 10\% . From this point forward, the DW results include the extrapolation below thresholds and the contribution from all initial autoionizing starting levels, including 2- to 3- L-shell hole transitions (MR1+MR2), unless clearly stated otherwise.

\begin{figure}
 \includegraphics[width=\columnwidth]{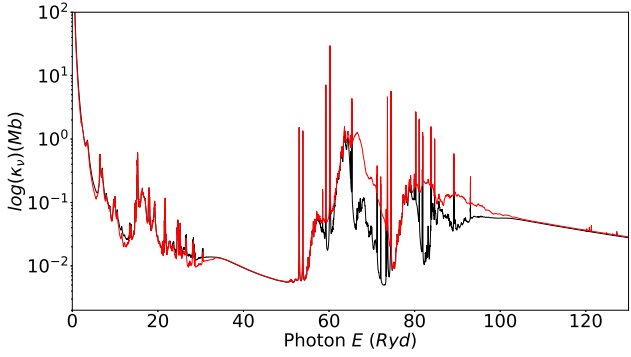}
\caption{Comparison between the DW monochromatic opacities using MR1 $n=6$ all starting levels (black) and the extended calculation (MR1+MR2) including two-to-three $L$-shell hole transitions (red).}
    \label{fig:Fig9}
\end{figure}

\subsubsection{Broadening}

As mentioned previously, the advantage of DW over $R$-matrix is the ability to collisionally broaden the Rydberg resonances in the same manner as for bound-bound lines. This may have a significant contribution to the mean opacities. Indeed, as seen in Figure 10, the numerous resonances give rise to broad features and enhance the Rosseland mean opacity by up to a factor of 2. 

\begin{figure}
 \includegraphics[width=\columnwidth]{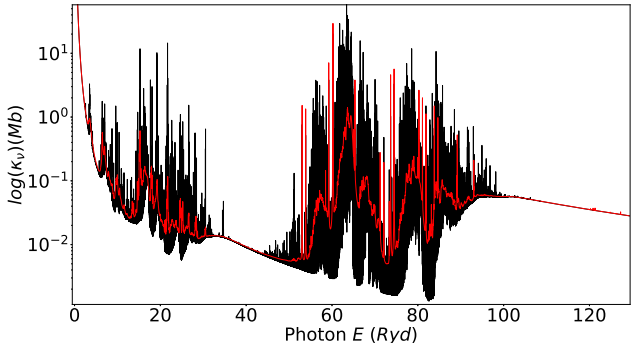}
\caption{Comparison between the DW monochromatic cross sections for MR1+MR2 $n=6$ without broadening of the resonances (black) and with broadening (red).}
    \label{fig:Fig10}
\end{figure}

\subsection{Comparison: DW, R-matrix and others}

\subsubsection{Comparison DW vs $R$-matrix}

In Figure 11 we compare our results using both methods (DW and $R$-matrix) for $n=6$, including all initial levels in the DW calculation and with the DW calculation for $n=6$ including only the true bound levels. First, we can see that the $R$-matrix $n=6$ calculation is more complete than the DW $n=6$ one without autoionizing starting states. (Recall, DW has less true bound levels.) This is especially so in the energy region of maximum importance for the Rosseland mean opacity that is to say $E\in[30-55$ Ryd$]$. The broadening of the resonances in the $R$-matrix data would raise the continuum in this region as it occurs in the case of DW data shown in the previous section. Otherwise, we see the general features are present in all the results and the behaviour is very similar.  When comparing with the DW results obtained with all the autoionizing starting levels, including the 2--3 holes ones, we see the importance of taking them into account. The low energy region, $E_{photon}<55$~Ryd, is dominated by these levels which make a strong contribution to the photoionization cross sections. As expected, the net effect is more than a factor 2 difference in the Rosseland means ($\kappa^{DW}_{R}=2.7337\times10^{-3}$,$\kappa^{R-matrix}_{R}=1.2834\times10^{-3}$.)
Another difference comes from the different structure, which translates into a small shift of the broad features when DW is compared to $R$-matrix.

The lack of broadening of the resonances in the $R$-matrix data increases the disagreement in the low energy region, while for energies greater than $50$~Ryd we see some large features and expect very similar curves when the resonances start to blend with some broadening, as seen in the previous section. We may expect that the implementation of such broadening will ensure that the two methods come into closer agreement at low energy as well. Finally, it must also be added that the $R$-matrix data can benefit from an extension of starting levels, including some of the autoionizing levels by a simple extrapolation of the existing cross-sections along identified Rydberg series. We did this exercise and end up with an increase in the Rosseland mean of around 5\%.

\begin{figure}
 \includegraphics[width=\columnwidth]{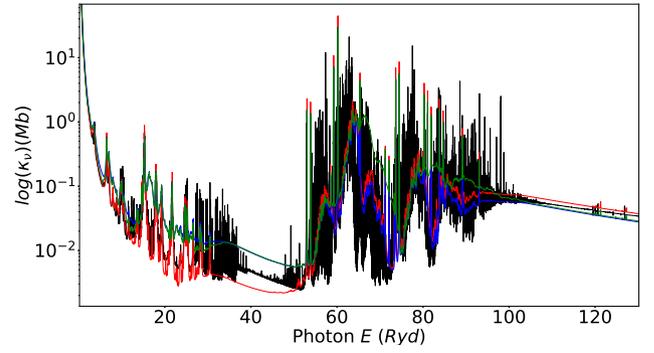}
\caption{Comparison of monochromatic opacities for Fe~XVII between $R$-matrix $n=6$ (black), DW MR1 $n=6$ with only true bound levels (red), DW MR1 $n=6$ with autoionizing starting levels (blue) and DW  MR1+MR2 $n=6$ plus 2- to 3- L-shell holes transitions (green).
}
    \label{fig:Fig11}
\end{figure}

However, the broadening for these features is very delicate. Indeed, how realistic is the model of broadening, which is a transposition of line broadening to the resonances, presented in \cite{Nahar2016a}? How should the plasma effects modify the interference between the direct and indirect process of photoionization? Is a simple broadening of the Fano's profiles sufficient? Is the profile more deeply affected and is the interference giving rise to another kind of profiles and shifts? While we are not yet able to answer these questions, the solution adopted to include broadening in a calculation will be yet another approximation pending possible improvement.

\subsubsection{Comparison with OPCD and SNAKP}

OP2005 and the new $R$-matrix results are the closest in terms of treatment. As expected, in Figure 12, the black (new $R$-matrix) and the blue (OP2005) lines are very similar. The new $n=6$ result shows more resonances, as expected, while the OP2005 has a higher continuum in the $[30-55$~Ryd$]$ region. This is due to the inclusion of the inner-shell in the OP2005 using a DW treatment and, hence, allowing for the broadening of the resonances in that region.  
We would expect a rise in the continuum when broadening is implemented for the resonances directly in the $R$-matrix data.

It is important to note that the differences in the Fe~XVII mean opacities compared to OP2005 (with inner shells) are lower by 34\%, 32\% and 31\% when using new  $R$-matrix Models A, B and C,  respectively. This difference is slightly different from the expected differences (35\%, 33\% and 31\%). This may come from the presence of inner shells in OP2005 which are not totally accounted for in the new data and even less as the maximum $n$ in the configuration set decreases. 

When comparing the previous release from OP \cite{Badnell2005} to the new results from the AS $n=6$ calculation including autoionizing starting levels, we found the same general features at energies greater then 55 Ryd. But at lower energies the inclusion of new levels, especially the autoionizing levels generates the large increase in the continuum. Of course a direct implication is the large enhancement of the Rosseland mean opacities of Fe~XVII. The increase reaches 55\% for the full $n=5$ case, with autoionizing levels allowing for 2--3 hole transitions. In contrast, the Planck mean opacity is 24\% lower. 

\begin{figure}
 \includegraphics[width=\columnwidth]{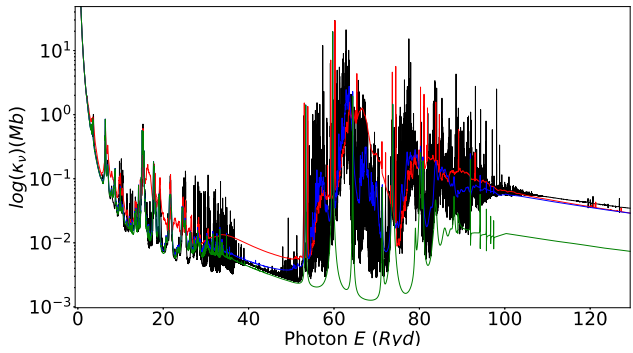}
\caption{Comparison between DW monochromatic opacities: MR1+MR2 $n=6$ all levels (red), $R$-matrix $n=6$ (black), OPCD 2005 (release 2 from OP, blue) and OPCD 1996 (release 1 from OP, green) }
    \label{fig:Fig12}
\end{figure}

We now proceed with considering the effect of the plasma broadening
in the work of Nahar and Pradhan \cite{Nahar2016a} (SNAKP hereafter). 
The SNAKP data presented below are taken from the plot by \cite{Nahar2016a} using the freeware g3data  (http://github.com/pn2200/g3data).  We can see in Figure 13 the effect of the broadening used by \cite{Nahar2016b} which uses the data from their Figure 5. When shown on the same graph we see it appears to show a somewhat unusual behaviour in the monochromatic cross sections. Indeed, around $70$~Ryd, while there is a dip in their cross sections as in ours and no significant resonances, the broadened ones are enhanced by one order of magnitude. This cannot come from the two strong lines seen after (at $75 $~Ryd) since they are bound-bound transitions and hence already broadened in our treatment. In the high-energy region, the broadened opacities give rise to a continuum below the non-broadened ones. These features show that some artefacts are due to the algorithm used for their broadening, with a possible impact on the final results. While the broadening issue is not the cause of the differences with the experimental data, as previously mentioned in \cite{Blancard2016}, it is certainly crucial in the calculation of monochromatic and mean opacities. It raises the value of $\kappa_{R}$, as we have shown previously.

\begin{figure}
 \includegraphics[width=\columnwidth]{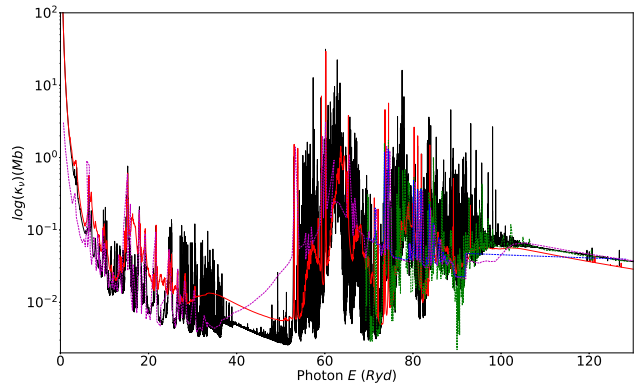}
\caption{Comparison between DW monochromatic opacities for MR1+MR2 $n=6$ all levels (red), $R$-matrix $n=6$ (black), SNAKP2016 with broadening (dashed - blue) and SNAKP2016 without broadening (dashed - green) and Pradhan \& Nahar 2018 (dashed - magenta) }
    \label{fig:Fig13}
\end{figure}

We also examined the latest result from \cite{Pradhan18} where they studied the accuracy and completeness of their calculation on Fe~XVII. One striking difference comes from the large enhancement of the monochromatic cross sections in the region of main contribution to the Rosseland mean opacities $[35-55$ Ryd$]$. Indeed, in our $R$-matrix calculations going up to the $n=6$ complex and where the resolution has been carefully studied, as well as in our DW calculation including the autoionizing levels as starting levels (more than 50000 levels) we could not see such a large increase in this region where no large contribution from resonances is expected from our results. In Figure 13, the magenta dashed line shows the same drop as SNAK where there are resonances in their calculation,  around $75$~Ryd and between $95$~Ryd and $100$~Ryd, but it shows a large contribution well above all our results in the $[35-55]$~Ryd region. We cannot understand this increase here, which looks like some unidentified broadening or other contributions. Of course, Nahar and Pradhan mention the contribution to the continuum of the extra 50,000 levels they included via a distorted-wave calculation, but this appears very different from any other similar DW calculations, including ours. At any rate, the large rise in the continuum in the  $[35-55]$~Ryd region is clearly the source of their large increase in the Rosseland mean. We cannot identify the source of this and may only speculate that is may be a re-normalisation problem for the population when incorporating their new DW data. We already presented the changes in the OP2005 release due to such approximations of re-normalisation. It was considered negligible (a few \%)  at that time for the applications. Today the inclusion of such a large number of levels changes the picture.  Unfortunately, even such a large increase does not solve the differences with the experimental data, as already pointed out by \cite{Blancard2016}. Moreover, while \cite{Bailey2015} presented the differences in the Rosseland mean opacity, the comparison only concerns a partial section of the energy range, i.e. the one for which experimental data are available. This implies an integration for the range $E \in [70-130$ Ryd$]$. Now, when \cite{Pradhan18} claim in their Table 1 a ratio of 1.65 compared to OP2005, it seems to apply to the Rosseland mean opacity of Fe~XVII over the total energy range and not just the range covered by the experiment. So, the comparisons in \cite{Bailey2015} and \cite{Pradhan18} do not appear to be strictly equivalent.

\subsubsection{Total Fe opacity and Comparison with experimental data}

In order to compare with the experimental data it is necessary to perform the same kind of improved calculations for all important ions present in the plasma for the given conditions. In the present case $T_e=2.1\times 10^6$~K and $N_e=3.1\times 10^{22}$~cm$^{-3}$. The main ions present in such Fe plasma are Fe~XVI, Fe~XVII, Fe~XVIII, Fe~XIX and Fe~XX. New atomic data calculations are underway for these ions. However, a first analysis can be performed using the new data for Fe~XVII in combination with the the older OP2005 data for the other ions. In Figure 14 we compare the monochromatic opacities for Fe with the our new $R$-matrix and DW data for Fe~XVII to the OP2005 results and the experimental data.

Given the results for Fe~XVII discussed in the previous sections and due to differences in the procedure to include the new Fe~XVII data to the total Fe mixture, we will analyse separately $R$-matrix and DW results. 
The complete comparisons for the Rosseland and Planck mean cross sections for Fe~XVII and Fe as well as the opacity means are presented in Table 2 to 4. In Table 2 (Planck cross sections) and Table 3 (Rosseland cross sections) we are presenting the results for the different Fe XVII model alongside the values for the Fe when these new models are included in the total Fe calculation.

For $R$-matrix, while we obtained Fe~XVII mean opacities significantly lower for the new $R$-matrix data (-20\%, -32\% for $n=5$ and -5\%,-31\% for n=6 in $\kappa_P$ and $\kappa_R$ respectively, these differences being clearly understood as detailed in the previous sections, we also have differences in the total Fe mean opacities of (-5\%, -9\% for $n=5$ and -2\%, -9\% for $n=6$ in $\kappa_P$ and $\kappa_R$ respectively). In addition to all the effects presented in the previous sections due to the new data we must add the effect of the new structures for Fe~XVII onto the ionization fractions. As we can see in T
able 1 the ionic fractions strongly depend on the structure of Fe~XVII which differs between models. The differences in the ionization potentials are of the order of $2.5\%$ to $5\%$ for all ions but Fe~XVII between the three new $R$-matrix model (A:$n=4$, B:$n=5$ and C:$n=6$) and reach $10\%$ to $28\%$ for Fe~XVII. The new calculation always produces a lower Fe~XVII content and a higher fraction of the other ions compared to the OP2005 data. Hence, the changes in the cross sections and transitions probabilities of Fe~XVII per se have a reduced direct effect, but they also change the ionic fractions and hence indirectly change the total Fe opacities. 
When comparing the results in Table 2 and 3 between the $R$-matrix models and OP2005, we clearly see the effect of the modification of the ionic fraction from Table 1 due to the inclusion of the new Fe~VXII data. 
Finally, model C with the inclusion of  extrapolated cross sections and extra bound-bound transitions, shows a closer result in the ionic fraction and in the partial and total cross sections when compared to 2005 release.
While the $R$-matrix data has improved significantly with the present calculations, the net effect is not yet fully displayed since no broadening is included and other ions need also to be treated to extract the proper estimations.

With the DW Fe~XVII data, we used the original OP2005 ionic fractions to translate the net effect of the new atomic data into the means (Rosseland and Planck).

Comparing with the DW calculation for Fe~XVII, as expected we reach the largest difference. The main effect being the inclusion of all the autoionizing levels as starting levels as well as the global broadening of all resonances alongside all bound-bound lines. We have clearly shown in the previous sections the large increase of the continuum in the $[30-55]$~Ryd energy range which, as expected, enhances the Rosseland means greatly. Hence, while the models including only true bound starting levels, as in the $R$-matrix datasets, since they do not include the contribution up to $n=9$ as opposed to the OP2005 inner-shells which do, $\sigma_R$(Fe~XVII) are lower than the OP2005 release and are very similar to the $R$-matrix values.
Once the autoionizing starting levels are allowed, as some were already present in the OP2005 results, the sign of the difference changes directly, even with the $n=5$ DW model. The differences are $\sim 14$\% (MR1 $n=5$ with auto), $\sim $44\% (MR1 $n=6$ with auto) and $\sim 65$\% (MR1+MR2 $n=6$ auto plus 2--3 holes) as seen in Tables 2 and 3 for Fe~XVII when compared to OP2005. This then translates directly into an increase in the total Fe Rosseland mean of 13\%. On the contrary, the Planck mean is reduced for the reason discussed previously due to the combination of dilution and height lowering in the lines and resonances which have been broadened in the present DW Fe~XVII models.  

Finally, in Table 4 we compare our total Fe Planck and Rosseland mean opacities to two other theoretical models. This comparison highlights some of the results shown previously. However, it has to be taken with caution since only one ion among the five main contributors has been updated. The Planck mean opacities derived from $R$-matrix data, while close to OP2005,  are 10\% higher than SCRAM results (S. Hansen private communication, using the code SCRAM \cite{Hansen2007}) and 10\% lower than the OPAS results (C. Blancard private communication, using OPAS  \cite{Blancard2012}). 
These results are certainly emphasizing the importance of broadening and suggest that a distinction between the treatment reserved for the resonance and the bound-bound transitions would be welcome. 
For the Rosseland mean opacities the present results show the signature of the limitation of the exercise since new calculations for all other ions will affect the results more drastically than for just a single ion. However, it is interesting to see the strong signature of the presence of the autoionizing starting levels, indeed, we reach a 15\% increase in the present DW results compared to the OP2005. But they are still much lower compared to the two other theoretical results.
Of course we cannot anticipate the final results on Fe when all data for the other ions will be included since many compensation and cancellation effects  will occur. However, we expect some increase in the total Fe Rosseland mean opacities and maybe a decrease in the Planck opacities, pending verification as intuition does not replace computations in cases as complex as the ones studied here. Indeed a new fully DW calculation will imply a complete broadening of the lines and resonances. In the $R$-matrix results a different treatment needs to be applied to the resonances compared to the bound-bound transitions. 
A careful build-up and implementation of the broadening treatment is crucial since this effect affects Planck and Rosseland opacities in different ways. 

\begin {table}
\caption {Ionic fractions as a function of Fe~XVII models (at $T_e=2.1\times 10^6$~K and $N_e=3.1\times 10^{22}$~cm$^{-3}$).} \label{tab:IonPop}
\begin{tabular}{ l l l l l }
   Fe~XVII Model & Fe~XX & Fe~XIX & Fe~XVIII & Fe~XVII \\
   \hline
   OPCD 2005 & 0.09759 &  0.28416  & 0.37238  & 0.19583   \\
   \hline
   $R$-matrix &  0.10273 & 0.29912   & 0.39199   & 0.15349  \\
    $n=$5 with 132 levels & & & & \\
    \hline
   $R$-matrix & 0.10127 &  0.29488 & 0.38643  &  0.16549  \\
    $n=6$ with 172 levels & & & & \\
    \hline
   $R$-matrix & 0.099859  & 0.29077  & 0.38104  & 0.17713   \\
    $n=6$ with 327 levels & & & & \\
    \hline
   $R$-matrix & 0.099898  & 0.29089 & 0.38119  & 0.17680   \\
    $n=6$ with 358 levels &  &  & &\\
    \hline
 \end{tabular}
 \end{table}
 
 \begin {table}
\caption {Planck mean cross sections in $a_0^2$ (at $T_e=2.1\times 10^6$~K and $N_e=3.1\times 10^{22}$~cm$^{-3}$) for Fe~XVII and  Fe, for various Fe~XVII models.} \label{tab:Plank}
\begin{tabular}{ l l l  }
   Fe~XVII Model & $\sigma_P$ (Fe~XVII) & $\sigma_P$(Fe) \\
   \hline
   OPCD 2005 & $1.4077\times 10^{-2}$ & $1.1290\times 10^{-2}$   \\
   $R$-matrix $n=5$ with 132 levels & $1.1187\times 10^{-2}$  & $1.1407\times 10^{-2}$  \\
   $R$-matrix $n=6$ with 172 levels & $1.3472\times 10^{-2}$  & $ 1.1085\times 10^{-2}$ \\
   $R$-matrix $n=6$ with 327 levels & $1.2697\times 10^{-2}$  & $1.0963\times 10^{-2}$ \\
   $R$-matrix $n=6$ with 358 levels & $1.2876\times 10^{-2}$  & $1.1012\times 10^{-2}$ \\
   AS-DW MR1 $n=5$ true bound initial& 1.2446$\times 10^{-2}$  & - \\
   AS-DW MR1 $n=6$ true bound initial& 1.2571$\times 10^{-2}$  &  - \\
   AS-DW MR1 $n=5$ +auto initial & $9.3728\times 10^{-3}$  &  -  \\
   AS-DW MR1 $n=6$ +auto initial & $8.8703\times 10^{-3}$  &  -  \\
   AS-DW MR1$+$MR2 $n=6$ all initial & $1.5784\times 10^{-2}$  &  $1.1130\times 10^{-2}$ \\
   \\
 \hline
 \end{tabular}
 \quad \newline

\end{table}

\begin {table}
\caption {Rosseland mean cross sections in $a_0^2$ (at $T_e=2.1\times 10^6$~K and $N_e=3.1\times 10^{22}$~cm$^{-3}$) for Fe~XVII and  Fe, for various Fe~XVII models.} \label{tab:Rosseland}
\begin{tabular}{ l l l }
   Fe~XVII Model & $\sigma_R$ (Fe~XVII) & $\sigma_R$(Fe) \\
   \hline
   OPCD 2005 & $1.9069\times 10^{-3}$ & $1.4356\times 10^{-3}$   \\
   $R$-matrix $n=5$ with 132 levels &$1.3453\times 10^{-3}$  & $1.3581\times 10^{-3}$  \\
   $R$-matrix $n=6$ with 172 levels & $1.3764\times 10^{-3}$  & $1.3561\times 10^{-3}$ \\
   $R$-matrix $n=6$ with 327 levels & $1.4559\times 10^{-3}$  & $1.3588\times 10^{-3}$ \\
   $R$-matrix $n=6$ with 358 levels & $1.5026\times 10^{-3}$  & $1.3686\times 10^{-3}$ \\
   AS-DW MR1 $n=5$ true bound initial& $1.5524\times 10^{-3}$  & - \\
   AS-DW MR1 $n=6$ true bound initial& $1.2213\times 10^{-3}$  & - \\
   AS-DW MR1 $n=5$ +auto initial & $2.6528\times 10^{-3}$  &  - \\
   AS-DW MR1 $n=6$ +auto initial & $2.7567\times 10^{-3}$  &  - \\
   AS-DW MR1+MR2 $n=6$ all initial & 3.1514$\times 10^{-3}$  & $1.6254\times 10^{-3}$ \\
   \\
    \hline
 \end{tabular}
 \quad \newline
\end{table}

\begin {table}
\caption {Fe Planck and Rosseland mean opacities (in cm$^2$/g) (at $T_e=2.1\times 10^6$~K and $N_e=3.1\times 10^{22}$~cm$^{-3}$), for various Fe~XVII models.} \label{tab:Fe means}
\begin{tabular}{ l l l }
   Fe~XVII Model & $\kappa_P$ (Fe) & $\kappa_R$(Fe) \\
   \hline
   OPCD 1996 & 1258 & 220.5 \\
   OPCD 2005 & 3497 & 460.7 \\
   $R$-matrix $n=5$ with 132 levels & 3445 & 409.7  \\
   $R$-matrix $n=6$ with 172 levels & 3567  & 411.8\\
   $R$-matrix $n=6$ with 327 levels & 3547 & 412.8 \\
   $R$-matrix $n=6$ with 358 levels & 3561  & 415.6 \\
   AS-DW MR1+MR2 $n=6$ with all initial & 3361 &  491.3 \\
   OPAS (private communication) & 3775 & 759.6 \\ 
   SCRAM (private communication) &  3110 & 878.0
   \\
    \hline
 \end{tabular}
 \quad \newline
Note: Our present results may change significantly either way (up or down) if we consider more neighbouring ion stages of Fe (see text for details).

\end{table}

As another attempt to estimate the impact of our present Fe~XVII calculation on the final Fe monochromatic opacities, we compare it to the revised experimental data \citep{Nagayama2019} as well as to the 2015 results \citep{Bailey2015}. The first thing we may expect is that the continuum at high energy is not going to increase with any of the new  Fe~XVII models. As seen above, any extension of targets does not affect significantly the results in the energy range above 100 Ryd where SANDIA's experiment is showing a much larger value. It has to be mentioned that this discrepancy has been reduced by half since the revision of the analysis of the experimental data in \cite{Nagayama2019}.  Another aspect of the discrepancy comes from the 'windows' filled in the experimental data but not in theoretical calculations.  It is important to note that the addition of the 2- to 3-holes autoionizing levels in the DW Fe~XVII models is clearly producing a contribution within the energy range considered in the experiment. In Figure 14 we have plotted the monochromatic opacities for a pure Fe~XVII plasma alongside the experimental data. The opacities present a clear increase in the 80 Ryd. and 90 Ryd (2 of the 3 so called 'windows') when the 2- to 3-holes autoionizing levels are taken into account. Of course one has to be cautious. The effect of population and ionic fraction will certainly temper the results. The final conclusion will come with the other ions which may or may not exhibit the same features.   

In Figure 15 we present the total Fe opacities obtained by replacing the OP2005 Fe~XVII data with the present results. As expected, the impact of the new data are not reducing substantially the discrepancies at high energy. While the continuum part seems hard to change, the filling of the windows remains a challenge and requires new data for the other ions concerned, as seen in Figure 16, where the new data for Fe~XVII are beginning to fill some gaps.  As already mentioned, the Rosseland and the Planck means are definitely affected and will certainly be more-so when all ions of interest have been included.

\begin{figure}
 \includegraphics[width=\columnwidth]{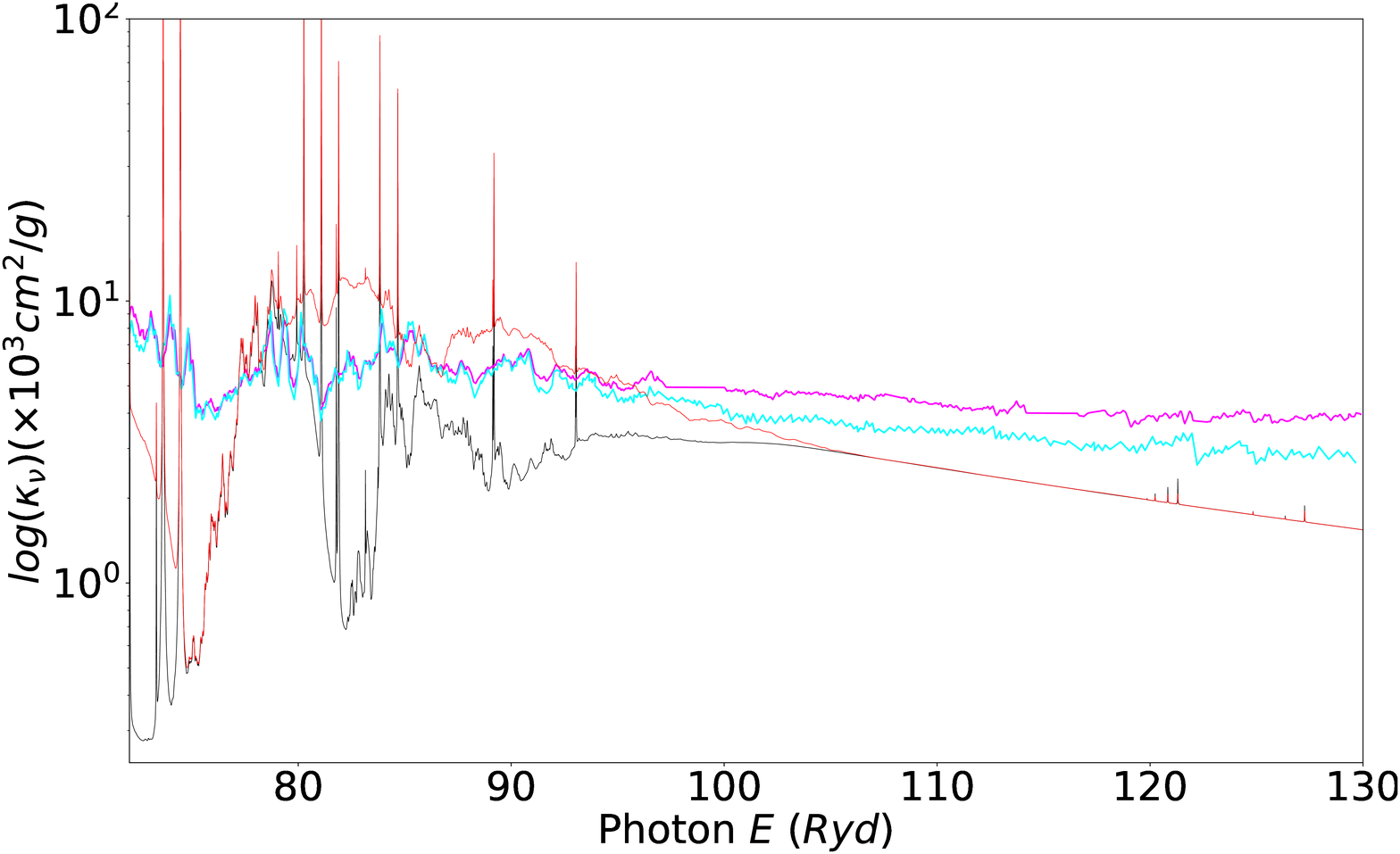}
\caption{ Fe~ XVII monochromatic opacities for pure Fe~XVII: DW $n=6$ with autoionizing starting levels: MR1 (black), MR1$+$MR2 (red) both compared to SANDIA's team experimental results from Bailey et al. (2015)(magenta) and their revision from Nagayama et al. (2019) (cyan).}
    \label{fig:Fig14}
\end{figure}

\begin{figure}
 \includegraphics[width=\columnwidth]{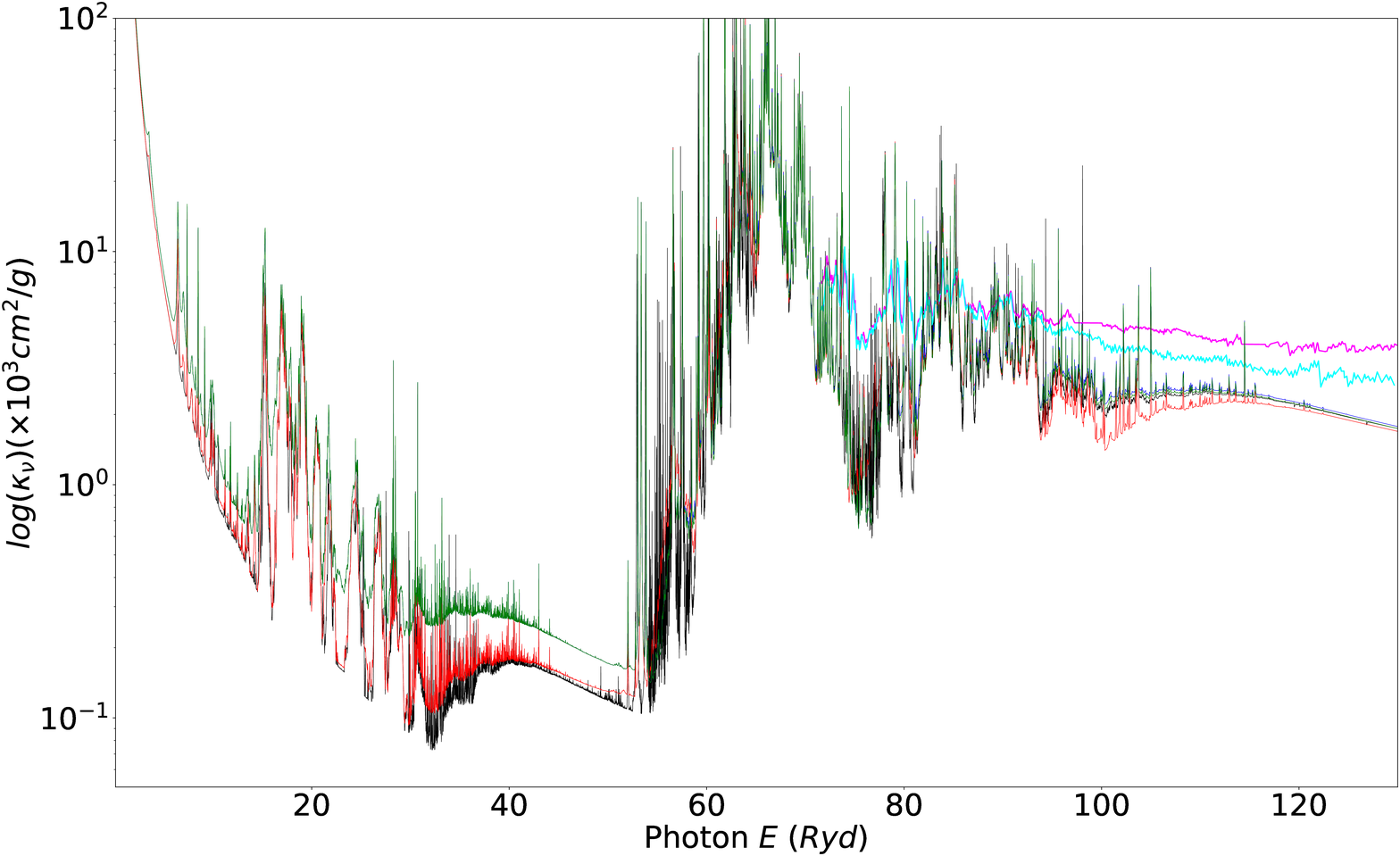}
\caption{ Fe opacities with new radiative data. DW for Fe~XVII $n=6$ with autoionizing starting levels: MR1 (blue), MR1$+$MR2 (green); $R$-matrix $n=6$ data (red) for Fe~XVII; both compared to OP 2005 release (black) and SANDIA's team experimental 
results from Bailey et al. (2015)(magenta) and their revision from Nagayama et al. (2019) (cyan). }
    \label{fig:Fig15}
\end{figure}

\begin{figure}
\includegraphics[width=\columnwidth]{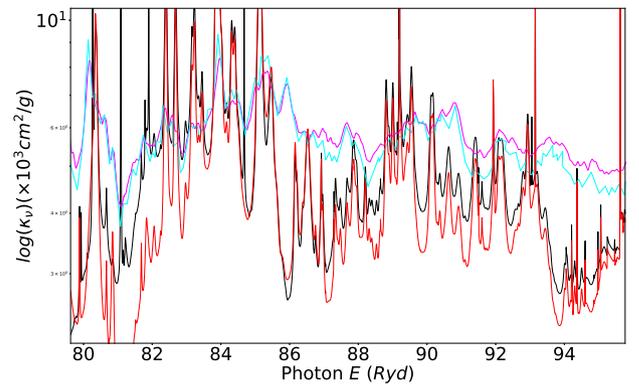}
\caption{ Fe opacities with new radiative data. DW for Fe~XVII $n=6$ with autoionizing starting levels: MR1$+$MR2 (black); OP 2005 release (red) and SANDIA's team experimental 
results from Bailey et al. (2015)(magenta) and their revision from Nagayama et al. (2019) (cyan).}
    \label{fig:Fig16}
\end{figure}

\section{Summary and Conclusions}


We have carried-out large-scale $R$-matrix and DW calculations of Fe~XVII radiative data to generate monochromatic opacities. We have thoroughly studied the impact of the different approximations on the final results and compared the methods. We have also compared them to previous calculations and experimental data.

The main results are summarized in Tables 2-4 of section 4.3.3 with Table 4 including the values for $\kappa_P$ and $\kappa_P$ from OPAS (CEA) and SCRAM (SANDIA/LANL).

We have tested different target expansions and  reached convergence in both sets of calculations while going from $n=4$ to $n=6$ for both methods. In the $R$-matrix calculation for Fe~XVII, the final results display a variation of less than 5\% in the Rosseland mean opacity between the $n=6$ and $n=4$  and 2\% $n=6$ and $n=5$ expansions. 

We have tested the limit of the $R$-matrix resolution and reached satisfactory results with 80000 points for the photoionization cross sections which do not change the Rosseland and Planck mean opacities by more than 0.2\% and 2\%,
respectively.

While the $R$-matrix method automatically interacts all of the resonances arising from the given target expansion, 
unlike the DW approach, the lack of collisional resonance broadening is the limiting factor in the accuracy of the results presented here.
While SNAKP have reported broadened $R$-matrix results, they give no details of their methodology and we have 
seen (section 4.3.2) that it appears to give rise to some unusual behaviour in the monochromatic opacities.
In contrast, the DW method allows us to treat each resonance as a bound-bound line which can be broadened at will. 

In the DW approach, we were able to extend the calculations further, including the contribution from initial autoionizing levels with two L-shell holes (which is not yet possible computationally in the $R$-matrix approach). The two-to-three L-shell hole transitions clearly 
make a significant contribution to the Rosseland and Planck mean opacities that cannot be neglected. The autoionizing levels as a whole contribute for more than 50\% to the Rosseland mean opacities of Fe~XVII with a 10\% increase coming from the two-to-three L-shell hole transitions.  

Overall, the net effect on the Fe Rosseland mean opacity reaches an increase of ~7\%  at the test $(T_e, N_e)$ when comparing the total opacities using the new data for Fe~XVII (DW $n=6$ MR1+MR2) compared to the OP release from 2005. Our new $R$-matrix Fe~XVII data, while missing resonance broadening and autoionizing initial levels, produce Rosseland mean opacities ~10\% below the results from OP2005. The comparison with the data from other groups (OPAS-CEA, SCRAM-SANDIA) shows the importance of the new treatment of the atomic data in the results from The Opacity Project. When considering the Planck mean opacities the new results sit just in between the other theoretical ones. This clearly indicates the different sensitivities to different aspects  of the calculations of the two different means. It is important to determine clear constraints for each mean in order to define the proper global treatment to calculate opacities. 
While the present Rosseland mean opacity results are still very low in comparison to other theoretical results, the present data indicate that our new calculations and those underway for the other relevant Fe ions may significantly reduce the difference.
(The focus in this paper has been on the physics describing the atomic processes rather than trying to converge the means.)  

As far as the comparison with the experiment is concerned, it clearly shows that nothing in the continuum at high energy has been improved. However, the work on the other ions concerned will be of the utmost importance. The presence of the autoionizing levels in the present work for Fe~XVII has already shown some features which fill some windows. If it is confirmed for the others ions we may partly resolve the existing discrepancy with experimental observations. But it remains to be established.

Both computational methods employed here (DW and $R$-matrix) are complementary and could be combined, as we already did in the OP2005 release, in order to increase accuracy and completeness at the same time.

$R$-matrix and DW calculations for other Fe ions  (Fe~XVIII, Fe~XIX, Fe~XX) are underway with the aim of giving us a complete picture of the global effect on the Fe opacity under the conditions at the base of the solar convection zone.  The interplay between all the different effects does not allow us a reliable intuitive prediction of the results that will be obtained from these new larger calculations. For example, one of the strong effects turns out to be the change in the ionic fraction introduced by the change of the structure in each particular ion.

It was found previously that the OP results were lower than OPAL (or OPAS) for some Fe-peak elements present in the Solar mixture and higher in others (C,N,O ...) but the Rosseland mean opacities for the solar mixture agreed  to within less than 4\%. Should our Fe results be followed by a large revision of the stellar Fe opacities, a global revision would become necessary. Its consequences would affect more than just stellar opacities. It is important to pave the way to the next generation of stellar OP opacities and astrophysical opacities. The next large set of calculations for all elements must be prepared and the different issues viz. ($R$-matrix) broadening, resonances and autoionizing level contributions must be better analyzed in any attempt for an improved treatment.   

\section*{Data Availability}
The data underlying this article considered as the best effort for Fe~XVII will be available on the Opacity Project public database Topbase and through the associated services accessible at http://cdsweb.u-strasbg.fr/topbase/testop/home.html .

\section*{Acknowledgements}
We would like to thank Stephanie Hansen (SANDIA) and Christophe Blancard (CEA) for sharing their data. FD would like to thank Claude Zeippen for fruitful discussions and remarks on the present paper. The work of FD was supported by the "Programme National de Physique Stellaire" (PNPS) of CNRS/INSU co-funded by CEA and CNES. FD would also like to thank the "Action Fédératrice Etoile" (AFE) funded by the Observatoire de Paris, for support.
NRB was supported by UK STFC UK APAP Network consolidated grant ST/R000743/1 with the University of Strathclyde. CPB would like to thank the UK STFC consolidated grant
ST/P000312/1 awarded to Queen's University of Belfast.




\bibliography{Delahaye_Fe17}

\begin{thebibliography}{27}
\providecommand{\natexlab}[1]{#1}
\providecommand{\url}[1]{\texttt{#1}}
\expandafter\ifx\csname urlstyle\endcsname\relax
  \providecommand{\doi}[1]{doi: #1}\else
  \providecommand{\doi}{doi: \begingroup \urlstyle{rm}\Url}\fi

\bibitem[{Asplund} et~al.(2004){Asplund}, {Grevesse}, {Sauval}, {Allende
  Prieto}, and {Kiselman}]{Asplund2004}
M.~{Asplund}, N.~{Grevesse}, A.~J. {Sauval}, C.~{Allende Prieto}, and
  D.~{Kiselman}.
\newblock {Line formation in solar granulation. IV. [O I], O I and OH lines and
  the photospheric O abundance}.
\newblock \emph{\aap}, 417:\penalty0 751--768, Apr. 2004.
\newblock \doi{10.1051/0004-6361:20034328}.

\bibitem[{Asplund} et~al.(2005){Asplund}, {Grevesse}, and
  {Sauval}]{Asplund2005}
M.~{Asplund}, N.~{Grevesse}, and A.~J. {Sauval}.
\newblock {The Solar Chemical Composition}.
\newblock In T.~G. {Barnes}, III and F.~N. {Bash}, editors, \emph{Cosmic
  Abundances as Records of Stellar Evolution and Nucleosynthesis}, volume 336
  of \emph{Astronomical Society of the Pacific Conference Series}, page~25,
  Sept. 2005.

\bibitem[{Asplund} et~al.(2009){Asplund}, {Grevesse}, {Sauval}, and
  {Scott}]{Asplund2009}
M.~{Asplund}, N.~{Grevesse}, A.~J. {Sauval}, and P.~{Scott}.
\newblock {The Chemical Composition of the Sun}.
\newblock \emph{\araa}, 47\penalty0 (1):\penalty0 481--522, Sept. 2009.
\newblock \doi{10.1146/annurev.astro.46.060407.145222}.

\bibitem[Badnell(2011)]{Badnell2011}
N.~R. Badnell.
\newblock {A Breit-Pauli distorted wave implementation for AUTOSTRUCTURE}.
\newblock \emph{Computer Physics Communications}, 182:\penalty0 1528--1535,
  2011.

\bibitem[Badnell and Seaton(2003)]{BadnellSeaton2003}
N.~R. Badnell and M.~J. Seaton.
\newblock {On the importance of inner-shell transitions for opacity
  calculations}.
\newblock \emph{Journal of Physics B Atomic Molecular Physics}, 36:\penalty0
  4367--4385, Nov. 2003.

\bibitem[{Badnell} et~al.(2005){Badnell}, {Bautista}, {Butler}, {Delahaye},
  {Mendoza}, {Palmeri}, {Zeippen}, and {Seaton}]{Badnell2005}
N.~R. {Badnell}, M.~A. {Bautista}, K.~{Butler}, F.~{Delahaye}, C.~{Mendoza},
  P.~{Palmeri}, C.~J. {Zeippen}, and M.~J. {Seaton}.
\newblock {Updated opacities from the Opacity Project}.
\newblock \emph{\mnras}, 360:\penalty0 458--464, June 2005.
\newblock \doi{10.1111/j.1365-2966.2005.08991.x}.

\bibitem[{Bailey} et~al.(2015){Bailey}, {Nagayama}, {Loisel}, {Rochau},
  {Blancard}, {Colgan}, {Cosse}, {Faussurier}, {Fontes}, {Gilleron},
  {Golovkin}, {Hansen}, {Iglesias}, {Kilcrease}, {Macfarlane}, {Mancini},
  {Nahar}, {Orban}, {Pain}, {Pradhan}, {Sherrill}, and {Wilson}]{Bailey2015}
J.~E. {Bailey}, T.~{Nagayama}, G.~P. {Loisel}, G.~A. {Rochau}, C.~{Blancard},
  J.~{Colgan}, P.~{Cosse}, G.~{Faussurier}, C.~J. {Fontes}, F.~{Gilleron},
  I.~{Golovkin}, S.~B. {Hansen}, C.~A. {Iglesias}, D.~P. {Kilcrease}, J.~J.
  {Macfarlane}, R.~C. {Mancini}, S.~N. {Nahar}, C.~{Orban}, J.-C. {Pain}, A.~K.
  {Pradhan}, M.~{Sherrill}, and B.~G. {Wilson}.
\newblock {A higher-than-predicted measurement of iron opacity at solar
  interior temperatures}.
\newblock \emph{\nat}, 517:\penalty0 56--59, Jan. 2015.
\newblock \doi{10.1038/nature14048}.

\bibitem[{Ballance} and {Griffin}(2006)]{Ballance2006}
C.~P. {Ballance} and D.~G. {Griffin}.
\newblock {Relativistic radiatively damped R-matrix calculation of the
  electron-impact excitation of W46+}.
\newblock \emph{J. Phys. B, At. Mol. Phys.}, 39\penalty0 (17):\penalty0 3617,
  Aug. 2006.
\newblock \doi{10.1088/0953-4075/39/17/017}.

\bibitem[Blancard et~al.(2012)Blancard, Coss\'e, and Faussurier]{Blancard2012}
C.~Blancard, P.~Coss\'e, and G.~Faussurier.
\newblock {Solar Mixture Opacity Calculations Using Detailed Configuration and
  Level Accounting Treatments}.
\newblock \emph{\apj}, 745:\penalty0 10, Jan. 2012.

\bibitem[{Blancard} et~al.(2016){Blancard}, {Colgan}, {Coss{\'e}},
  {Faussurier}, {Fontes}, {Gilleron}, {Golovkin}, {Hansen}, {Iglesias},
  {Kilcrease}, {MacFarlane}, {More}, {Pain}, {Sherrill}, and
  {Wilson}]{Blancard2016}
C.~{Blancard}, J.~{Colgan}, P.~{Coss{\'e}}, G.~{Faussurier}, C.~J. {Fontes},
  F.~{Gilleron}, I.~{Golovkin}, S.~B. {Hansen}, C.~A. {Iglesias}, D.~P.
  {Kilcrease}, J.~J. {MacFarlane}, R.~M. {More}, J.-C. {Pain}, M.~{Sherrill},
  and B.~G. {Wilson}.
\newblock {Comment on ``Large Enhancement in High-Energy Photoionization of Fe
  XVII and Missing Continuum Plasma Opacity''}.
\newblock \emph{Physical Review Letters}, 117\penalty0 (24):\penalty0 249501,
  Dec. 2016.
\newblock \doi{10.1103/PhysRevLett.117.249501}.

\bibitem[{Colgan} et~al.(2013{\natexlab{a}}){Colgan}, {Kilcrease}, {Magee},
  {Armstrong}, {Abdallah}, {Sherrill}, {Fontes}, {Zhang}, and
  {Hakel}]{Colgan2013b}
J.~{Colgan}, D.~P. {Kilcrease}, N.~H. {Magee}, G.~S.~J. {Armstrong},
  J.~{Abdallah}, M.~E. {Sherrill}, C.~J. {Fontes}, H.~L. {Zhang}, and
  P.~{Hakel}.
\newblock {Light element opacities from ATOMIC}.
\newblock \emph{High Energy Density Physics}, 9:\penalty0 369--374, June
  2013{\natexlab{a}}.
\newblock \doi{10.1016/j.hedp.2013.03.001}.

\bibitem[{Colgan} et~al.(2013{\natexlab{b}}){Colgan}, {Kilcrease}, {Magee},
  {Armstrong}, {Abdallah}, {Sherrill}, {Fontes}, {Zhang}, and
  {Hakel}]{Colgan2013a}
J.~{Colgan}, D.~P. {Kilcrease}, N.~H. {Magee}, Jr., G.~S.~J. {Armstrong},
  J.~{Abdallah}, Jr., M.~E. {Sherrill}, C.~J. {Fontes}, H.~L. {Zhang}, and
  P.~{Hakel}.
\newblock {Light element opacities of astrophysical interest from ATOMIC}.
\newblock In J.~D. {Gillaspy}, W.~L. {Wiese}, and Y.~A. {Podpaly}, editors,
  \emph{American Institute of Physics Conference Series}, volume 1545 of
  \emph{American Institute of Physics Conference Series}, pages 17--26, July
  2013{\natexlab{b}}.
\newblock \doi{10.1063/1.4815837}.

\bibitem[{Colgan} et~al.(2015){Colgan}, {Kilcrease}, {Magee}, {Abdallah},
  {Sherrill}, {Fontes}, {Hakel}, and {Zhang}]{Colgan2015}
J.~{Colgan}, D.~P. {Kilcrease}, N.~H. {Magee}, J.~{Abdallah}, M.~E. {Sherrill},
  C.~J. {Fontes}, P.~{Hakel}, and H.~L. {Zhang}.
\newblock {Light element opacities of astrophysical interest from ATOMIC}.
\newblock \emph{High Energy Density Physics}, 14:\penalty0 33--37, Mar. 2015.
\newblock \doi{10.1016/j.hedp.2015.02.006}.

\bibitem[{Colgan} et~al.(2016){Colgan}, {Kilcrease}, {Magee}, {Sherrill},
  {Abdallah}, {Hakel}, {Fontes}, {Guzik}, and {Mussack}]{Colgan2016}
J.~{Colgan}, D.~P. {Kilcrease}, N.~H. {Magee}, M.~E. {Sherrill}, J.~{Abdallah},
  Jr., P.~{Hakel}, C.~J. {Fontes}, J.~A. {Guzik}, and K.~A. {Mussack}.
\newblock {A New Generation of Los Alamos Opacity Tables}.
\newblock \emph{\apj}, 817:\penalty0 116, Feb. 2016.
\newblock \doi{10.3847/0004-637X/817/2/116}.

\bibitem[{Delahaye} et~al.(2013){Delahaye}, {Palmeri}, {Quinet}, and
  {Zeippen}]{Delahaye2013}
F.~{Delahaye}, P.~{Palmeri}, P.~{Quinet}, and C.~J. {Zeippen}.
\newblock {IPOPv2: Photoionization of Ni XIV - a test case}.
\newblock In G.~{Alecian}, Y.~{Lebreton}, O.~{Richard}, and G.~{Vauclair},
  editors, \emph{EAS Publications Series}, volume~63 of \emph{EAS Publications
  Series}, pages 321--330, Dec. 2013.
\newblock \doi{10.1051/eas/1363036}.

\bibitem[{Delahaye} et~al.(2016){Delahaye}, {Zw{\"o}lf}, {Zeippen}, and
  {Mendoza}]{Delahaye2016}
F.~{Delahaye}, C.~M. {Zw{\"o}lf}, C.~J. {Zeippen}, and C.~{Mendoza}.
\newblock {IPOPv2 online service for the generation of opacity tables}.
\newblock \emph{JQSRT}, 171:\penalty0 66--72, Mar. 2016.
\newblock \doi{10.1016/j.jqsrt.2015.11.010}.

\bibitem[{Grant} et~al.(1980){Grant}, {McKenzie}, {Norrington}, and
  {Mayers}]{Grant1980}
I.~P. {Grant}, B.~J. {McKenzie}, P.~H. {Norrington}, and D.~F. {Mayers}.
\newblock {An Atomic Multiconfigurational Dirac-Fock Package}.
\newblock \emph{Comp, Phys. Comm.}, 21\penalty0 (2):\penalty0 271--231, Nov.
  1980.
\newblock \doi{10.1016/0010-4655(80)90041-7}.

\bibitem[{Hansen} et~al.(2007){Hansen}, {Bauche}, {Bauche-Arnoult}, and
  {Gu}]{Hansen2007}
S.~B. {Hansen}, J.~{Bauche}, C.~{Bauche-Arnoult}, and M.~F. {Gu}.
\newblock {Hybrid atomic models for spectroscopic plasma diagnostics}.
\newblock \emph{High Energy Density Physics}, 3\penalty0 (1-2):\penalty0
  109--114, May 2007.
\newblock \doi{10.1016/j.hedp.2007.02.032}.

\bibitem[{Mondet} et~al.(2015){Mondet}, {Blancard}, {Coss{\'e}}, and
  {Faussurier}]{Mondet2015}
G.~{Mondet}, C.~{Blancard}, P.~{Coss{\'e}}, and G.~{Faussurier}.
\newblock {Opacity Calculations for Solar Mixtures}.
\newblock \emph{\apjs}, 220:\penalty0 2, Sept. 2015.
\newblock \doi{10.1088/0067-0049/220/1/2}.

\bibitem[{Nagayama} et~al.(2019){Nagayama}, {Bailey}, {Loisel}, {Dunham},
  {Rochau}, {Blancard}, {Colgan}, {Coss{\'e}}, {Faussurier}, {Fontes},
  {Gilleron}, {Hansen}, {Iglesias}, {Golovkin}, {Kilcrease}, {MacFarlane},
  {Mancini}, {More}, {Orban}, {Pain}, {Sherrill}, and {Wilson}]{Nagayama2019}
T.~{Nagayama}, J.~E. {Bailey}, G.~P. {Loisel}, G.~S. {Dunham}, G.~A. {Rochau},
  C.~{Blancard}, J.~{Colgan}, P.~{Coss{\'e}}, G.~{Faussurier}, C.~J. {Fontes},
  F.~{Gilleron}, S.~B. {Hansen}, C.~A. {Iglesias}, I.~E. {Golovkin}, D.~P.
  {Kilcrease}, J.~J. {MacFarlane}, R.~C. {Mancini}, R.~M. {More}, C.~{Orban},
  J.~C. {Pain}, M.~E. {Sherrill}, and B.~G. {Wilson}.
\newblock {Systematic Study of L -Shell Opacity at Stellar Interior
  Temperatures}.
\newblock \emph{\prl}, 122\penalty0 (23):\penalty0 235001, June 2019.
\newblock \doi{10.1103/PhysRevLett.122.235001}.

\bibitem[{Nahar} and {Pradhan}(2016{\natexlab{a}})]{Nahar2016a}
S.~N. {Nahar} and A.~K. {Pradhan}.
\newblock {Large Enhancement in High-Energy Photoionization of Fe XVII and
  Missing Continuum Plasma Opacity}.
\newblock \emph{Physical Review Letters}, 116\penalty0 (23):\penalty0 235003,
  June 2016{\natexlab{a}}.
\newblock \doi{10.1103/PhysRevLett.116.235003}.

\bibitem[{Nahar} and {Pradhan}(2016{\natexlab{b}})]{Nahar2016b}
S.~N. {Nahar} and A.~K. {Pradhan}.
\newblock {Nahar and Pradhan Reply:}.
\newblock \emph{Physical Review Letters}, 117\penalty0 (24):\penalty0 249502,
  Dec. 2016{\natexlab{b}}.
\newblock \doi{10.1103/PhysRevLett.117.249502}.

\bibitem[{Norrington} and {Grant}(1987)]{Norrington1987}
P.~H. {Norrington} and I.~P. {Grant}.
\newblock {Low-energy electron scattering by Fe XXIII and Fe VII using the
  Dirac R-matrix method}.
\newblock \emph{J. Phys. B, At. Mol. Phys.}, 20\penalty0 (18):\penalty0
  4869--4881, Sept. 1987.
\newblock \doi{10.1088/0022-3700/20/18/023}.

\bibitem[{Pradhan} and {Nahar}(2018)]{Pradhan18}
A.~K. {Pradhan} and S.~N. {Nahar}.
\newblock {Recalculation of Astrophysical Opacities: Overview, Methodology, and
  Atomic Calculations}.
\newblock In \emph{Workshop on Astrophysical Opacities}, volume 515 of
  \emph{Astronomical Society of the Pacific Conference Series}, page~79, Aug.
  2018.

\bibitem[{Rogers} and {Iglesias}(1992)]{OPAL1992}
F.~J. {Rogers} and C.~A. {Iglesias}.
\newblock {Rosseland mean opacities for variable compositions}.
\newblock \emph{\apj}, 401:\penalty0 361--366, Dec. 1992.
\newblock \doi{10.1086/172066}.

\bibitem[{Seaton}(1987)]{ADOC_II}
M.~J. {Seaton}.
\newblock {Atomic data for opacity calculations. I - General description}.
\newblock \emph{Journal of Physics B Atomic Molecular Physics}, 20:\penalty0
  6363--6378, Dec. 1987.
\newblock \doi{10.1088/0022-3700/20/23/026}.

\bibitem[{Seaton} et~al.(1994){Seaton}, {Yan}, {Mihalas}, and
  {Pradhan}]{Seaton1994}
M.~J. {Seaton}, Y.~{Yan}, D.~{Mihalas}, and A.~K. {Pradhan}.
\newblock {Opacities for Stellar Envelopes}.
\newblock \emph{\mnras}, 266:\penalty0 805, Feb. 1994.
\newblock \doi{10.1093/mnras/266.4.805}.

\end{thebibliography}








\bsp	
\label{lastpage}
\end{document}